\documentclass[prb,twocolumn,amsmath,amssymb,superscriptaddress]{revtex4-2}
\usepackage{graphicx}
\usepackage{bm}

\def\be{\begin{equation}}
\def\ee{\end{equation}}
\def\bd{\begin{displaymath}}
\def\ed{\end{displaymath}}

\def\be{\begin{equation}}
\def\ee{\end{equation}}

\usepackage[hypertex]{hyperref}

\begin{document}

\thispagestyle{myheadings}

\title{Effects of exchange distortions in the magnetic Kagome lattice}

\author{A. A. Coker}
\affiliation{Department of Physics, University of North Florida, Jacksonville, FL 32224, USA}
\author{A. Saxena}
\affiliation{Theoretical Division and Center for Nonlinear Studies, Los Alamos National Laboratory, Los Alamos, New Mexico 87545, USA}
\author{J.T. Haraldsen}
\affiliation{Department of Physics, University of North Florida, Jacksonville, FL 32224, USA}



\date{\today}

\begin{abstract}

This study examines the effect of distorted triangular magnetic interactions  in the Kagome lattice. Using a Holstein-Primakoff expansion, we determine the analytical solutions for classical energies and the spin-wave modes for various magnetic configurations. By understanding the magnetic phase diagram, we characterize the changes in the spin waves and examine the spin distortions of the ferromagnetic (FM), Antiferrimagnetic (AfM), and 120$^{\circ}$ phases that are produced by variable exchange interactions and lead to various non-collinear phases, which provides a deeper understanding of the magnetic fingerprints of these configurations for experimental characterization and identification.

\end{abstract}

\maketitle

\section{Introduction}

\setkeys{Gin}{draft=false}
Complex magnetic lattice configurations have gained attention both theoretically and experimentally during the last decade as part of the search for topological and non-collinear magnetic orders, flat-band and Kitaev interactions, and the elusive quantum spin liquid state\cite{kita:06:AoP,knol:19:ARoCMP,gang:08:PRB,chis:15:PRL,hara:09:PRL,hara:09:JPCM,mukh:15:PRL,yang:20:PRR,zhou:17:RMP,sava:16:RPP,bale:10:nature,fish:18:book}. These magnetic systems are so interesting because of their potential in technology for sustainable, energy-efficient memory devices and computational power\cite{dien:11:HMM}.

\begin{figure} 
    \includegraphics[width=3.25in]{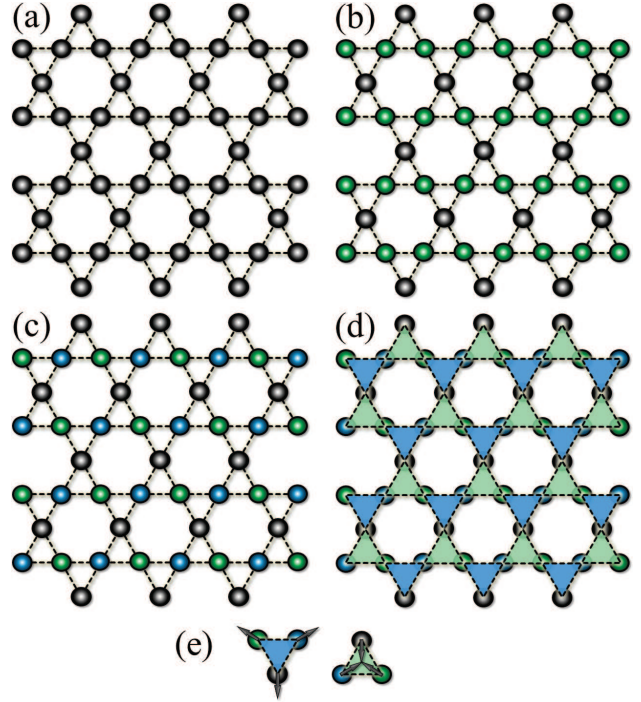}
    \caption{The atoms in a Kagome lattice are colored by spin orientation. The various spin configurations are the ferromagnetic (FM - all spin angles in the same direction) (a), antiferrimagnetic (AfM - spin angles are anti-aligned) (b), and the $120^\circ$ phase, where each spin is rotated 120 degrees away from its neighbors' (c). The $120^\circ$ phase makes up triangular ``in" and ``out" atomic spin groups (e), which further illustrate its AFM nature (d).}
    \label{structure}
\end{figure}

One such lattice that has increased in popularity is the Kagome lattice, which is described as a two-dimensional triangular three-sublattice structure and the foundational structure for the three-dimensional pyrochlore lattice\cite{essa:17:PRB,ross:09:PRL,gao:18:PRB,bent:20:arx}. This extra sublattice helps distinguish the Kagome from the two-sublattice honeycomb lattice\cite{boyk:18:PRB}. As shown in Fig. \ref{structure}(a), the Kagome lattice consists of a lattice of coupled trimers, which introduces multiple inversion points. Theoretical studies on materials with Kagome lattices have qualified their relevance to these technologies and have further probed for numerous magnetic and electronic properties\cite{wang:20:PCCP,harr:13:PRB,obri:10:PRB,li:18:SA,tran:20:arx,maks:17:PRB,ochi:17:JPSJ,schm:04:PRB,maks:15:PRB,yama:04:IJQC,moes:00:CJP,gov:02:JPCM,ower:16:PRB}. Experimental studies have offered a different perspective on many of the same properties\cite{zhan:20:PRB,shar:20:IEEE,wu:19:IEEE,yin:18:N,yin:19:NP,yazy:19:NP,rigo:07:PRL,zhan:20:PRB,yin:20:NC}. 

Previously, Boyko et al. examined the spin-wave dynamics of the Kagome lattice for three different magnetic configurations [out-of-plane ferromagnetic (FM), out-of-plane antiferrimagnetic (AfM), and 120$^{\circ}$ phase] and various isotropic nearest, next-nearest, and next-next-nearest neighbor interactions\cite{boyk:20:AdP}. In this work, it was shown that, to first order, the FM phase produced three modes, wherein two modes were dispersive, long-range order modes, and one was a non-dispersive, cluster-like flat band. These types of cluster modes are not unheard of as they have been observed in structures like the pyrochlore lattice\cite{gao:18:PRB}. Furthermore, Boyko et al. also revealed that, to first order, the 120$^{\circ}$ phase mimics the antiferromagnetic (AFM) honeycomb lattice due to the net in and out spin configurations\cite{boyk:20:AdP}. However, unlike the AFM honeycomb lattice, the Kagome lattice can break this degeneracy with second- and third-order interactions.

When considering the collinear Heisenberg model, exchange competition can produce frustration in the system and require an axial anisotropy to stabilize due to underlying non-collinear states\cite{boyk:20:AdP,hara:09:JPCM,fish:18:book}. Frustrated states can also come from the interactions between orbitals that typically result in FM or AFM orders but may become complicated by the competition between interactions as well as any induced crystal-field anisotropies. The frustration in the triangular interactions can lead to the need for more complex interactions to describe the excitations of the system, which is a fairly standard approach as any deviation of a known model indicates the need for new and exciting physics. In the Kagome lattice, this complexity tends to lead researchers into the realm of more exotic interactions like Kitaev model and spin liquids\cite{kita:06:AoP,gang:08:PRB,knol:19:ARoCMP}.

\begin{figure} 
    \includegraphics[width=3.25in]{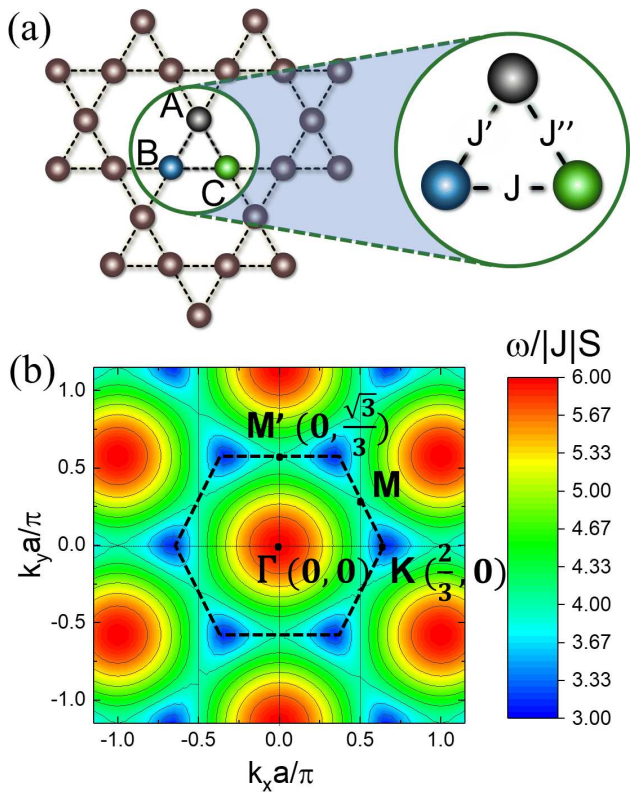}
    \caption{(a) The triangular subgeometry on which our calculations are focused. The exchange interactions $(J_{ij})$, which may not all be equal between all atom pairs.
    (b) A top-down view of the FM spin waves in a Kagome lattice with symmetry points M, K and $\Gamma$ overlaid.}
    \label{brillouin}
\end{figure}

Another avenue that can alleviate these frustrations is the presence of some underlying, possibly non-collinear, magnetic ground state that is not considered\cite{hara:09:PRL}. Given the large number of magnetic configurations for this structure, many groups are forced to examine numerical methods to interpret and understand neutron scattering experiments. While these numerical methods provide critical information for the characterization and identification of magnetic systems, there is little understanding of how the system's interactions compete to produce the given state, which can hinder further development of magnetic materials. By examining analytical solutions and the evolution of complexity for simple models, one can gain insight into the effects of exchange interactions on various configurations of spins in the Kagome lattice. These insights can aid experimentalists and theorists (or modelers) in the identification of different magnetic orders by simple comparison, which can, in turn, allow for the tunability of structures.

In this study, we look to understand many of the Kagome lattice's magnetic configurations by first producing analytical solutions for the spin waves of in-plane and out-of-plane ferromagnetic (FM) and antiferrimagnetic (AfM) arrangements along with the in-plane 120$^{\circ}$ phase, then numerically characterizing various spin states with a distorting nearest neighbor interaction. The term antiferrimagnetic denotes a mixed composition of AFM and FM aligned spins, which result in a structure that still produces a net magnetic moment (shown in Fig. \ref{structure}(b)). Using a Heisenberg spin-spin exchange Hamiltonian with on-site anisotropy, we determine the energy phase diagrams for this distorted system as well as the spin-wave dynamics. Furthermore, we look beyond the 120$^{\circ}$ phase and push the analytical limit by generalizing the in-plane magnetism to a 120$^{\circ}$ + $d\theta$ phase, where $d\theta$ goes from -120$^{\circ}$ (FM phase) to 60$^{\circ}$ (AfM phase).

These calculations allow for a detailed understanding of the changes in the spin-wave dynamics expected for various configurations of the magnetic Kagome lattice, which is useful for experimental identification and interpretation. Furthermore, deviation from these spin-wave dynamics provides evidence for non-standard interactions like those determined by the Kitaev model or other non-Heisenberg models.

\begin{figure}[!]
    \includegraphics[width=3.25in]{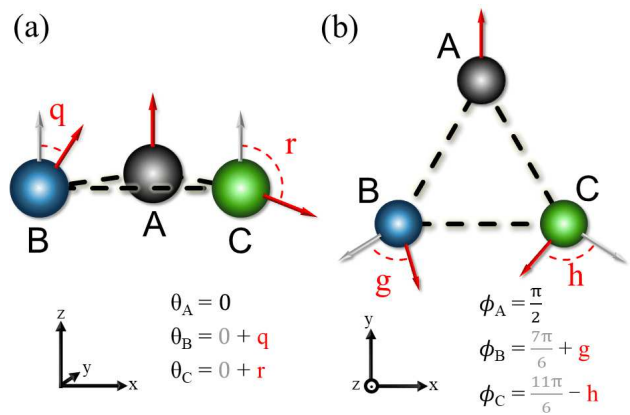}
    \caption{The spin angles for the three-atom structure are represented as deviations from a default state both in and out of the xy-plane. (a) The azimuthal spin angles $\theta$, illustrated by the red arrows on each atom, are represented as deviations from the angles $\theta_{FM}$ (gray arrows) that reflect the FM state. As the significance of these angles is their relation to one another rather than absolute orientation, the angle $\theta_A$ was set arbitrarily to 0. Deviation angles $q$ and $r$ are taken as the difference between angle $\theta$ and $\theta_{FM}$. (b) The in-plane spin angles $\phi$, illustrated by the red arrows on each atom, are represented as deviations from the angles $\phi_{120}$ (gray arrows) which make up the 120$^\circ$ phase. The angle $\phi_A$ is set as $\pi/2$. Deviation angles $g$ and $h$ are taken as the difference between angle $\phi$ and $\phi_{120}$}
    \label{GHandQRDiagram}
\end{figure}

\begin{figure}[!]
    \includegraphics[width=3.5in]{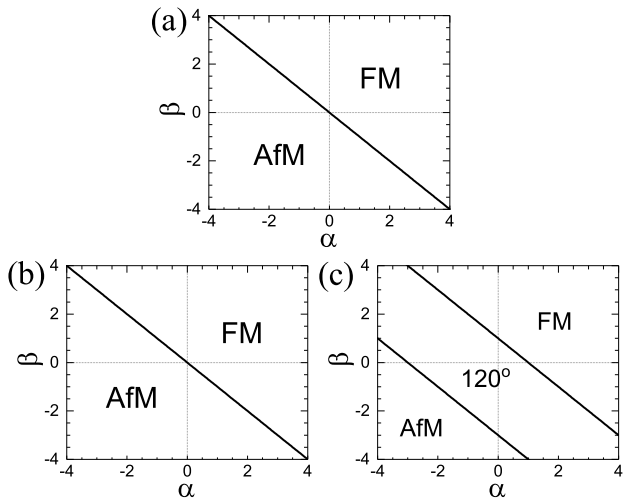}
    \caption{(a) Phase diagram for spin waves out of the plane for all values of $J$ and $\mathcal{D}$. (b) and (c) are the in-plane diagrams for all values of $\mathcal{D}$, where (b) is for positive $J$ values and (c) for negative values.  The phase borders were determined by setting the simplified classical energies for each pair of phases equal to each other. The configuration for each region was determined by testing which phase offered the lowest energy for the $\alpha$ and $\beta$ values encompassed by that region. There may be other spin configurations to consider in thoroughly characterizing these phases. However, these would require a considerable amount of extra parameterization. To remain within two dimensions and make as few assumptions as possible, we consider only the most basic configurations for this portion of the analysis.}
    \label{Phase_Diagrams}
\end{figure}

\section{Spin-exchange Hamiltonian}

To gain a complete understanding of how local interaction changes can affect the spin state of the Kagome lattice, a Heisenberg spin-spin Hamiltonian is used to model first-order interactions with variable exchange values (shown in Fig. \ref{brillouin}(a)). From this Hamiltonian, the energy eigenstates and spin-wave dynamics are examined within the analytical limit and expanded numerically to explore the spin evolution of the magnetic properties, where the spin-spin exchange Hamiltonian with $z$-axis anisotropy is given as
\begin{equation} \label{Energy2}
    H = -\frac{1}{2} \sum_{i \neq j} J_{ij} \bar S_i \cdot \bar S_j - D \sum_{i} \bar S_{iz}^2.
\end{equation}
Here, $J_{i,j}$ is the exchange interaction between the spin sites and $D$ is the anisotropy energy\cite{hara:09:JPCM,fish:18:book}. Since the Kagome lattice can produce numerous collinear and non-collinear phases, we must be able to consider the azimuthal and polar angles of the site spins. Therefore, the Hamiltonian must be shifted to study non-collinear spin configurations. As described in Ref. \cite{hara:09:JPCM}, the spin rotation is performed using an Euler rotation matrix $U$ upon the Hamiltonian, which is dependent on spherical coordinates $\theta$ (azimuthal) and $\phi$ (polar) between the two spins \cite{hara:09:JPCM,hara:09:PRL}. Applying this rotation to the Hamiltonian gives
\begin{equation} \label{Energy3}
    H = \displaystyle - \frac{1}{2} \sum_{i \neq j} J_{ij} \bar S_i \cdot \underline{U}_i\underline{U}_j^{-1} \bar S_j
    - D \sum_{i} \underline{U}_i^{-1} \bar S_{iz}^2.
\end{equation}

\begin{figure}
    \includegraphics[width=3.5in]{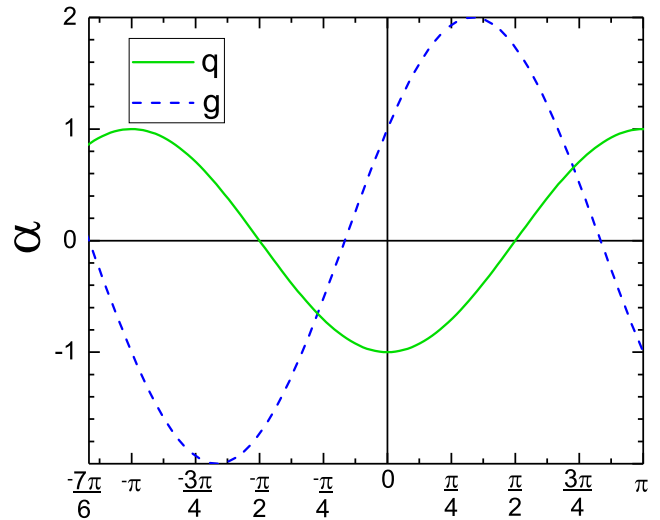}
    \caption{$\alpha$ as a function of $q$ and $g$ such that classical energy is minimized for the out-of-plane and in-plane spin configurations, respectively, where $J$ is positive. Negative $J$ values are characterized by a reflection of this graph about the $\alpha$ axis.}
    \label{AlphaVsQG}
\end{figure}

Through a $(1/S)$ Holstein-Primakoff expansion of this Hamiltonian, the system breaks into various orders 
\begin{equation} \label{hamexpansion}
    H = E_0 + H_1 + H_2 + \cdot \cdot \cdot .
\end{equation}
Here, $E_0$ is the classical energy, which can be used to determine the system's overall ground state for a given spin configuration. The $H_1$ term is the vacuum contribution to the spin dynamics, which vanishes in a stable system. $H_2$ produces the first-order contributions to the spin-wave dynamics within the quadratic limit. Higher order terms can also be determined. However, these terms produce quantum fluctuations which can be ignored for large $S$\cite{hara:09:JPCM}.

\begin{figure*}
    \includegraphics[width=\textwidth]{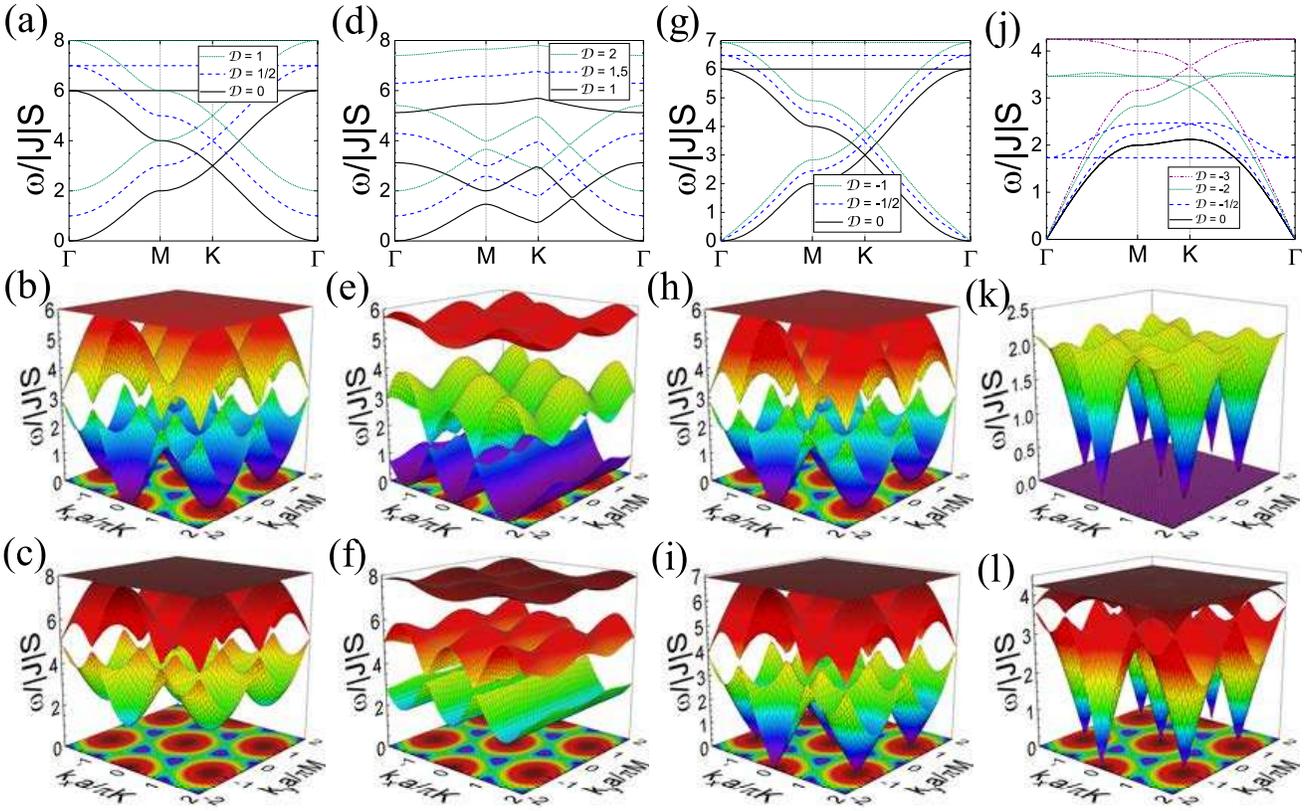}
    \caption{The spin-wave dynamics for the simplest spin configurations, where $J = J' = J''$ ($\alpha = \beta = \mathcal{J}$) and no magnetic field is present. (a), (d), (g), and (j) draw out the spin waves through the path in $k$-space which passes through symmetry points $\Gamma$, M, K, then back to $\Gamma$ for the out-of-plane FM, out-of-plane AfM, in-plane FM, and 120$^\circ$ configurations, respectively. (b) and (c) are the out-of-plane FM ($\mathcal{J}= 1$) spin waves with anisotropies $\mathcal{D}$ = 0 and $\mathcal{D}$ = 1, respectively. (e) and (f) show a similar evolution in the AfM configuration ($\mathcal{J} = -1$). However, since the equilateral spin waves are not stable without anisotropy, the anisotropies here are $\mathcal{D}$ = 1 (e) and $\mathcal{D}$ = 2 (f). (h) and (i) are the in-plane FM ($\mathcal{J} = 1$) spin waves with anisotropies $\mathcal{D}$ = 0 and $\mathcal{D}$ = -1, respectively. (k) and (l) are spin waves for the 120$^\circ$ ($\mathcal{J} = -1$) configuration. Although stable with none (k), (l) reflects a large anisotropy value of -3 to illustrate the full transformation that this configuration's spin waves undergo as this parameter is amplified. Color scales applied to the spin waves are grouped by configuration/column. All 3D figures include a projection of the central energy level of the FM configuration spin waves onto the $k$-plane to visualize the symmetry points. The heat map projection's colors are scaled relative to that of that single energy level's minima and maxima, as in figure \ref{brillouin}(b)}
    \label{Equilateral_Spinwaves}
\end{figure*}

The Kagome system can produce many spin configurations. The most well-known are collinear FM and AfM systems.  The spins in two configurations can be either in-plane or out-of-plane and make distinct changes in the spin dynamics. Outside of the collinear systems is the non-collinear 120$^{\circ}$ phase [shown in Figs. \ref{structure}(c)-(e)], which is typically an in-plane rotation of spin produced through AFM frustration.

First, we examine the classical energy for this system and then move on to the spin-dynamics. Within these sections, we will investigate the out-of-plane and in-plane structures for the FM and AfM phases. With those established, we will then enable a magnetic distortion of the exchange interactions and simulate the evolution of the spin dynamics of both collinear and non-collinear phases with various parameters.

\section{Classical Energy}

Before determining the spin-wave dynamics for specific configurations, the classical energy of the system is used to determine the ground-state spin configuration within the 3 sublattice (3-SL) Kagome system.

\begin{figure*}[!]
    \includegraphics[width=6.0in]{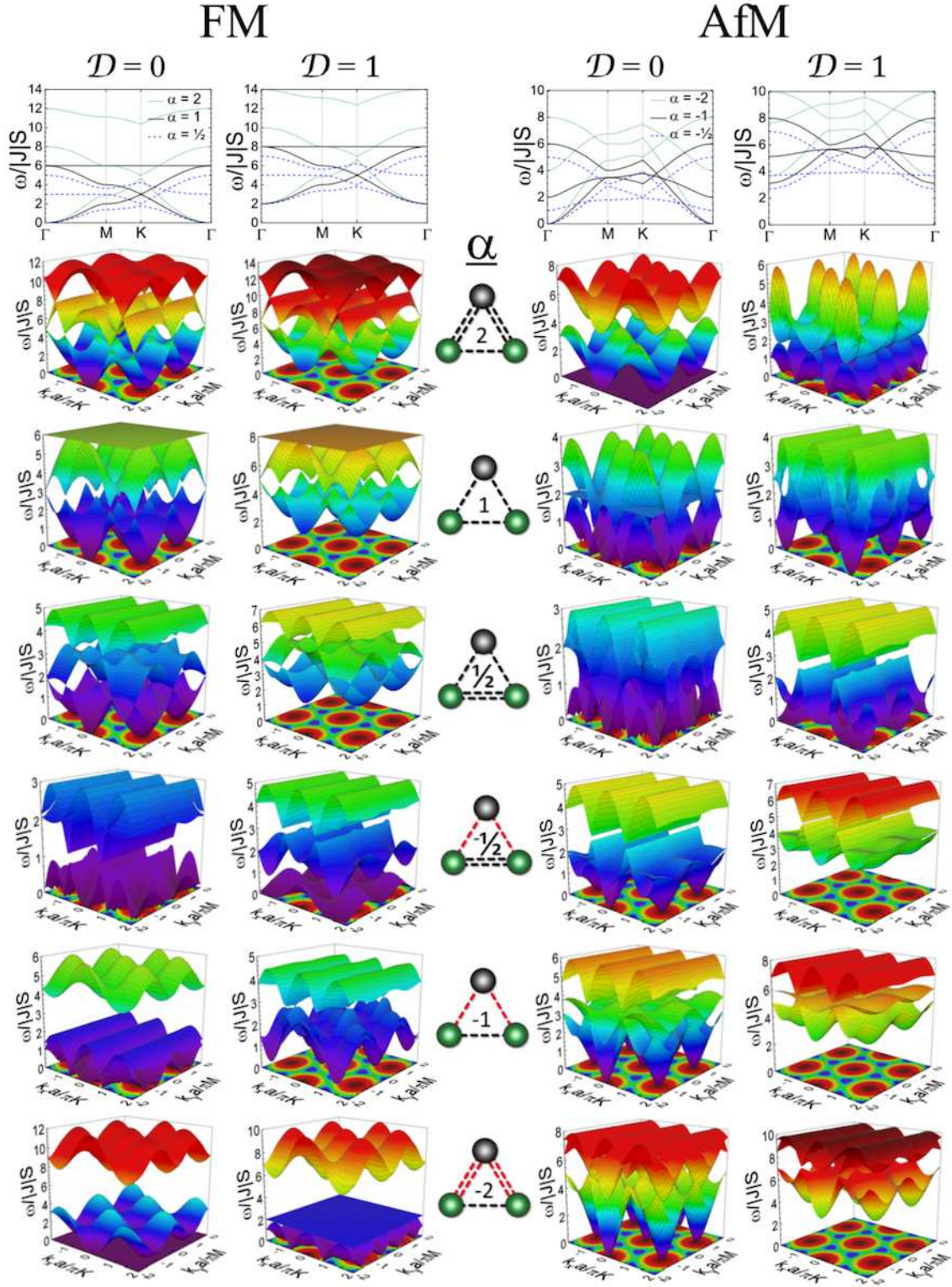}
    \caption{Variously distorted isosceles spin waves for the out-of-plane configurations where, in all cases, $\mathcal{J} = 1$. The first row shows the spin-wave energy values through the path in $k$-space that passes through symmetry points $\Gamma, M, K$, and back to $\Gamma$. These 2-D representations exclude half of the $\alpha$ values shown in the 3D graphs due to a large amount of overlap rendering the paths unreadable. The $\alpha$ value for each row of graphs is given by its central-column triangle diagram. The coloring of atoms indicates the AfM configuration. While the FM state would be more accurately by one uniform color, the dichromatic coloring of the AfM phase was chosen to represent either state here. Unstable spin waves were included to illustrate the results of the exchange interaction manipulations that oppose the intuited signs. For both configurations, the evolution of these graphs with anisotropy was included in its second column. All 3D FM graphs (first two columns) are color-grouped together and all 3D AfM graphs are similarly grouped.}
    \label{Isosceles_Spinwaves_Out}
\end{figure*}

\begin{figure*}[!]
    \includegraphics[width=6.0in]{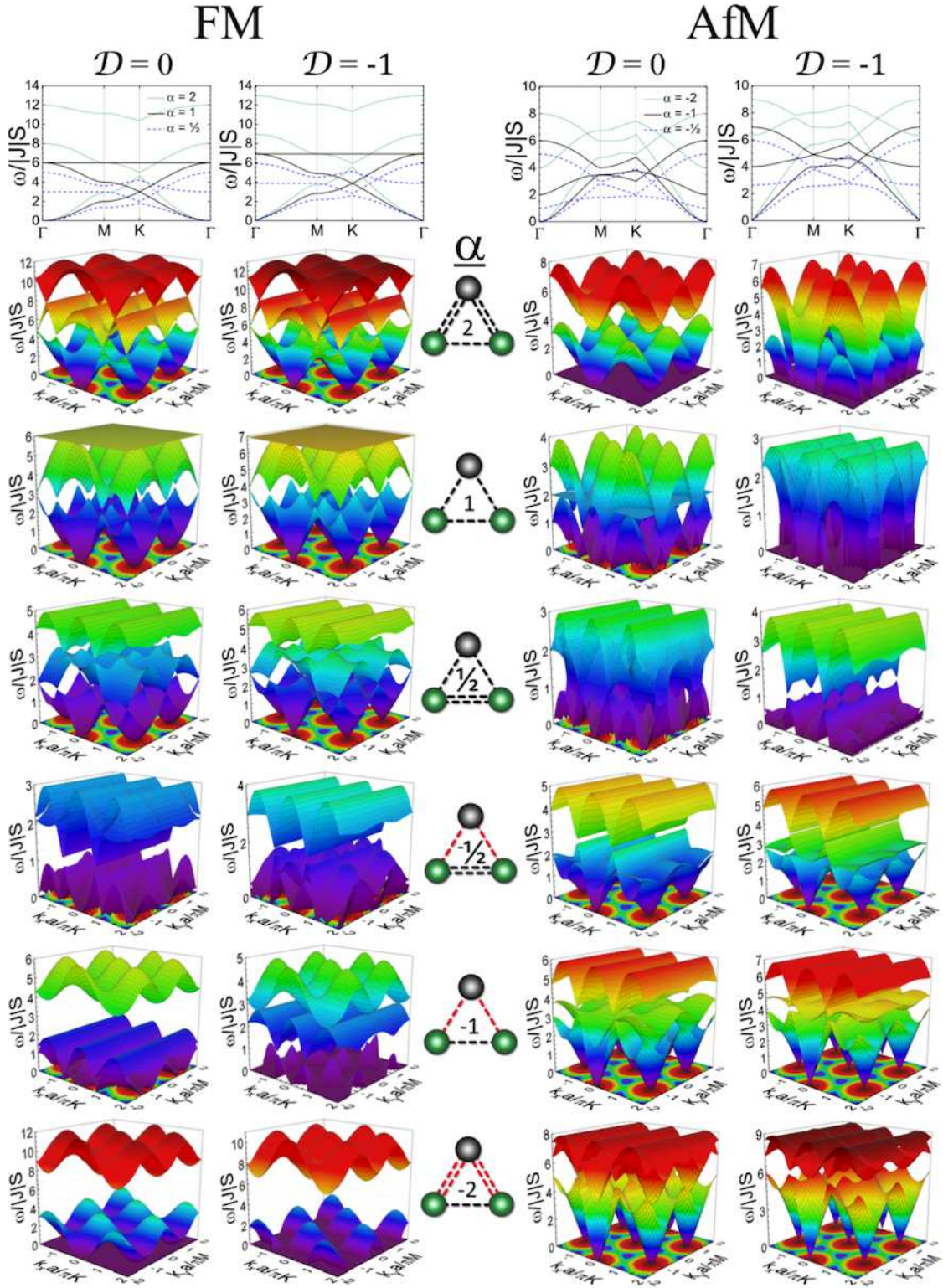}
    \caption{Variously distorted isosceles spin waves for the in-plane FM and AfM configurations, each of which has $\mathcal{J} = 1$. The first row shows the spin-wave energy values through the path in k-space that passes through symmetry points $\Gamma$, M, K, then back to $\Gamma$. These each exclude half of the $\alpha$ values shown in the 3D graphs, as a large amount of overlap with the other values made them too difficult to read. The $\alpha$ value for each row of graphs is given by the diagram in the central column. For both configurations, the evolution of the spin waves with anisotropy was included in the second column for each. FM graphs are color-grouped together and all AfM graphs are similarly grouped.}
    \label{Isosceles_Spinwaves_In}
\end{figure*}

The classical energy of each magnetic configuration is given by
\begin{equation} \label{E_0}
\begin{array}{ll}
    E_0 & \displaystyle = -\frac{1}{2}\sum_{i,j}J_{ij}S^2 \Big(\sin(\theta_i)\sin(\theta_j)\cos(\phi_j-\phi_i) \\
  & + \cos(\theta_i)\cos(\theta_j)\Big)
  \displaystyle - DS^2 \sum_i \cos^2(\theta_i),
\end{array}
\end{equation}
where the various spin angles are illustrated in Fig. \ref{GHandQRDiagram}\cite{hara:09:JPCM}. Within the different spin configurations, we will specialize angles for the purpose of achieving analytical and understandable solutions.

\subsection{Out-of-plane spin configurations}

\begin{figure*}[!]
    \includegraphics[width=7.05in]{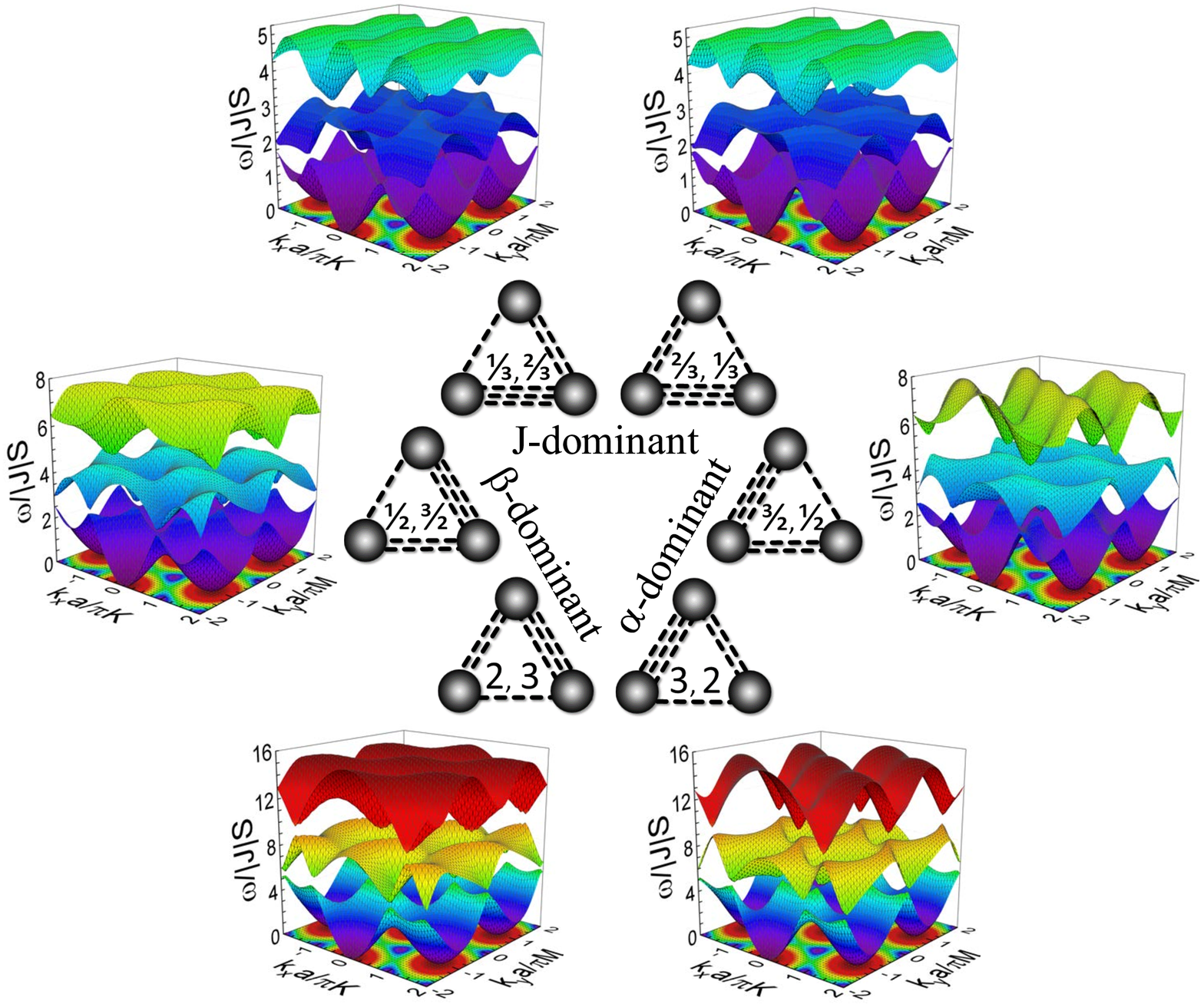}
    \caption{The FM spin waves configuration with scalene exchange interaction proportions. The $\alpha$ and $\beta$ values for each are separated by a comma as $\alpha$, $\beta$ in the corresponding triangle diagrams. All graphs are grouped together in a single color group.}
    \label{Scalene_Spinwaves_FM}
\end{figure*}

The out-of-plane spin configurations have collinear spins with $\theta$ = 0 or $\pi$. Here, the classical energy can be written as 
\begin{equation} \label{E_0_OUT}
\begin{array}{ll}
    \displaystyle \frac{E_0}{| J |S^2} = & -\frac{2}{3}\Big(\alpha\cos(\theta_A)\cos(\theta_B) + \beta\cos(\theta_A)\cos(\theta_C)  
    \\
    &\displaystyle +\mathcal{J}  \cos(\theta_B)\cos(\theta_C)\Big) - \mathcal{D},
\end{array}
\end{equation} 
where $\mathcal{J} \equiv J/| J | = \pm1$ and determines the general exchange interaction of the system. Additionally, $\mathcal{D} \equiv D/| J |$, $\alpha$ $\equiv$ $\frac{J'}{|J|}$ and $\beta$ $\equiv$ $\frac{J''}{|J|}$. By pulling out an overall $J$ and looking at the ratio of exchange interactions, we reduce the number of variables and produce an overall scaling factor that helps determine the energy scale of a material system.

For ease of calculation, angles $\theta_A$, $\theta_B$, and $\theta_C$ are replaced with the values 0, 0 + $q$, and 0 + $r$, respectively, where $q$ and $r$ are deviation angles from the FM state, as illustrated in Fig. \ref{GHandQRDiagram}(a). The energy becomes
\begin{equation} \label{E_0_OUT_2}
    \displaystyle \frac{E_0}{| J |S^2} = -\frac{2}{3}\Big(\alpha\cos(q)+\beta\cos(r) +  \mathcal{J}\cos(q)\cos(r)\Big) - \mathcal{D}.
\end{equation}
Assuming no magnetic field nor anisotropy, the energy function was inspected for minima and maxima. The values for $\alpha$ and $\beta$ that minimize $E_0$ out the plane belong to the sinusoids $\alpha = -\mathcal{J}\cos(r)$ and $\beta = -\mathcal{J}\cos(q)$. It is important to remember that this relationship is only accurate for $\theta$ values that are multiples of $\pi$, as the consideration of any other value reinstates the necessity for the first term of Eq. \ref{E_0}, leading these relationships to oscillate discretely between $-1$ and 1 as $(-1)^{n+1}\mathcal{J}$ where $n$ is $q/\pi$ for $\alpha$ and $r/\pi$ for $\beta$. 

The case where $\beta$ and $\alpha$ are equal to each other, but not necessarily to $J$, is termed the ``isosceles" case. While $r$ and $q$ are not mathematically required to be equivalent when $\alpha$ and $\beta$ are equal, it is likely that the angles will be equal through symmetry. The minimizing formula is then $\alpha = -\mathcal{J}\cos(q)$, which is shown in Fig. \ref{AlphaVsQG}. In order to minimize $E_0$, $J'$ may never exceed $J$ in magnitude. Assuming a positive $J$ value, the maximum value $J' = J$ requires the AfM configuration ($q$ = $\pi$) to minimize energy and the minimum value ($J' = -J$) requires the FM ($q$ = 0) configuration for minimization. 

\begin{figure*}[!]
    \includegraphics[width=7.0in]{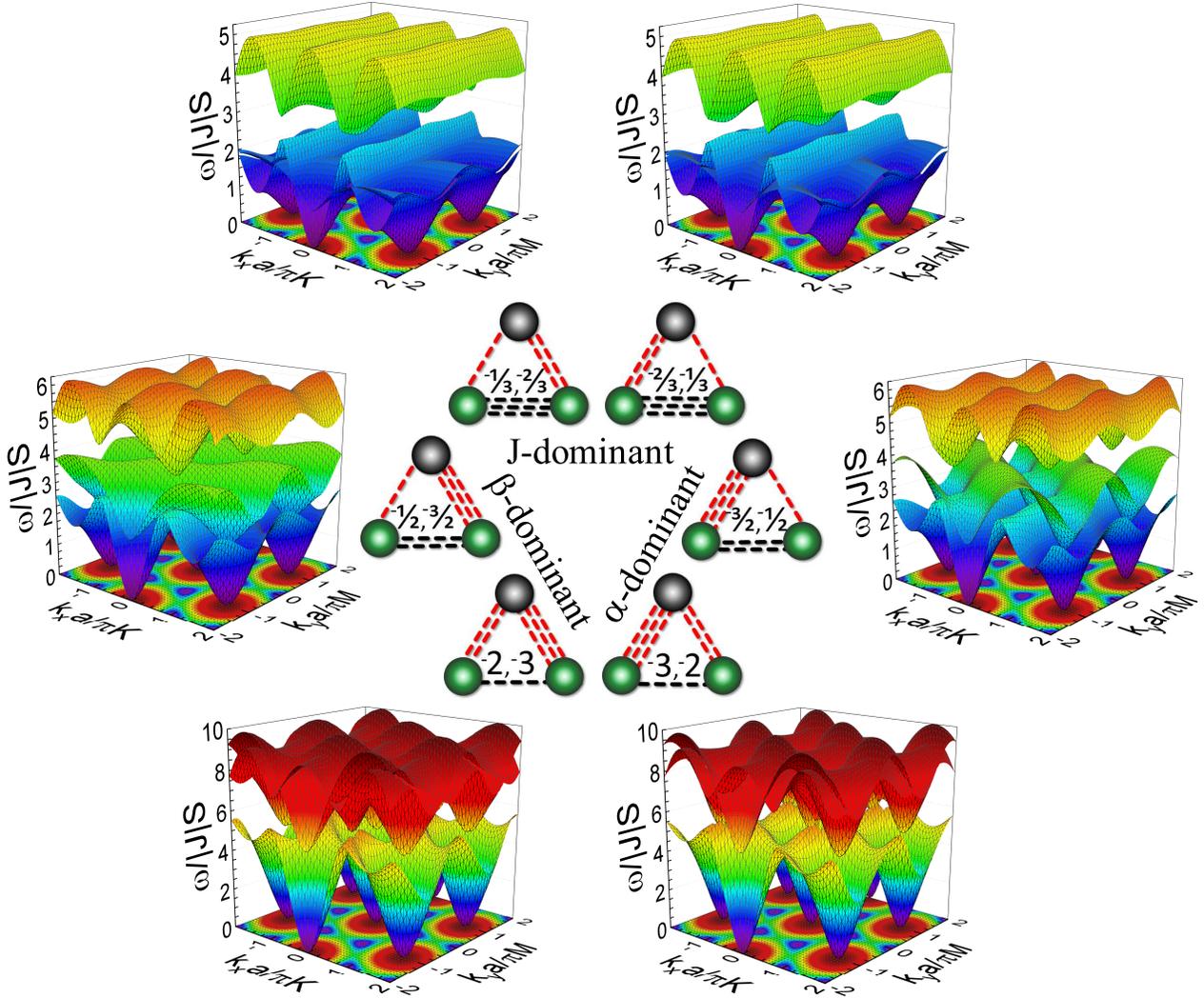}
    \caption{The spin waves for the AfM configuration with scalene exchange interaction proportions. The $\alpha$ and $\beta$ values for each case are separated by a comma as $\alpha, \beta$ in the corresponding triangle diagrams. A single color group encompasses all 3D graphs shown.}
    \label{Scalene_Spinwaves_AfM}
\end{figure*}

The simplest case, where all exchange interactions are equal ($\beta = \alpha = \mathcal{J}$), is termed the ``equilateral" case. In this state, the value for $g$ that minimizes $E_0/ | J |S^2$ is $\pi$, which represents the AfM configuration. $E_0/ | J |S^2$ is maximized by $g = 0$, indicating the FM configuration. These configurations are therefore the focus of our out-of-plane analyses.

The classical energy for the out-of-plane FM configuration ($\theta = 0$) is reduced to
\begin{equation} \label{E_OUT_FM}
    \displaystyle \frac{E_{0,FM}}{| J |S^2} = -\frac{2}{3}(\alpha+\beta + \mathcal{J})- \displaystyle \mathcal{D}
\end{equation}
and the AfM energy to 
\begin{equation} \label{E_OUT_AfM}
    \displaystyle \frac{E_{0,AfM}}{| J |S^2} = -\frac{2}{3}(-\alpha-\beta +   \mathcal{J})-\mathcal{D}.
\end{equation}

These energies were used to generate the phase diagram in Fig. \ref{Phase_Diagrams}(a).

\subsection{In-plane spin configurations}


For in-plane configurations ($\theta = \pi/2$), the classical energy becomes
\begin{equation} \label{E_0_IN}
\begin{array}{ll}
    \displaystyle \frac{E_0}{| J |S^2} = & \displaystyle -\frac{2}{3}\Big[\alpha \cos(\phi_A-\phi_B) + \beta \cos(\phi_A-\phi_C)\\ 
    & + \mathcal{J}\cos(\phi_B - \phi_C)\Big].
\end{array}
\end{equation}

To reduce the number of variables, the energy is rearranged in terms of deviation from the $120^{\circ}$ phase such that $\phi_A$, $\phi_B$, and $\phi_C$ become $(1/2)\pi$, $(7/6)\pi + g$, and $(11/6)\pi – h$, respectively, as shown in Fig. \ref{GHandQRDiagram}(b). Assuming all variables are real valued, the energy becomes
\begin{equation}\label{E_0_IN_Isosceles}
\begin{array}{ll}
    \displaystyle \frac{E_0}{| J |S^2}  = & \displaystyle
    \frac{2}{3}\Big[\textstyle\alpha\sin\Big(\frac{\pi}{6} + g\Big) + \beta\sin\Big(\frac{\pi}{6} + h\Big)\\
    & + \cos\Big(\textstyle\frac{\pi}{3} + g + h\Big)\Big].
\end{array}
\end{equation}
It is important to note that since we are using the isotropic Heisenberg model, all spins in this configuration can be rotated in the plane by any phase factor of $\phi^{\prime}$ as it is energetically degenerate. We choose this particular orientation to simplify the expressions. Using this system, the relationships between spin angle and exchange interaction strength which minimize the classical energy are determined analytically as:
\begin{equation}\label{alpha_beta_min_E_IN}
    \displaystyle \alpha = 
    \mathcal{J}\frac{\sin\Big(\frac{\pi}{3}+g+h\Big)}{\cos\Big(\frac{\pi}{6}+g\Big)} ~{\rm and}~
    \displaystyle \beta = 
    \mathcal{J}\frac{\sin\Big(\frac{\pi}{3}+g+h\Big)}{\cos\Big(\frac{\pi}{6}+h\Big)} .
\end{equation}
Unlike the out-of-plane case, these relationships hold true for any value of g and h.

\begin{figure*}[!]
    \includegraphics[width=\textwidth]{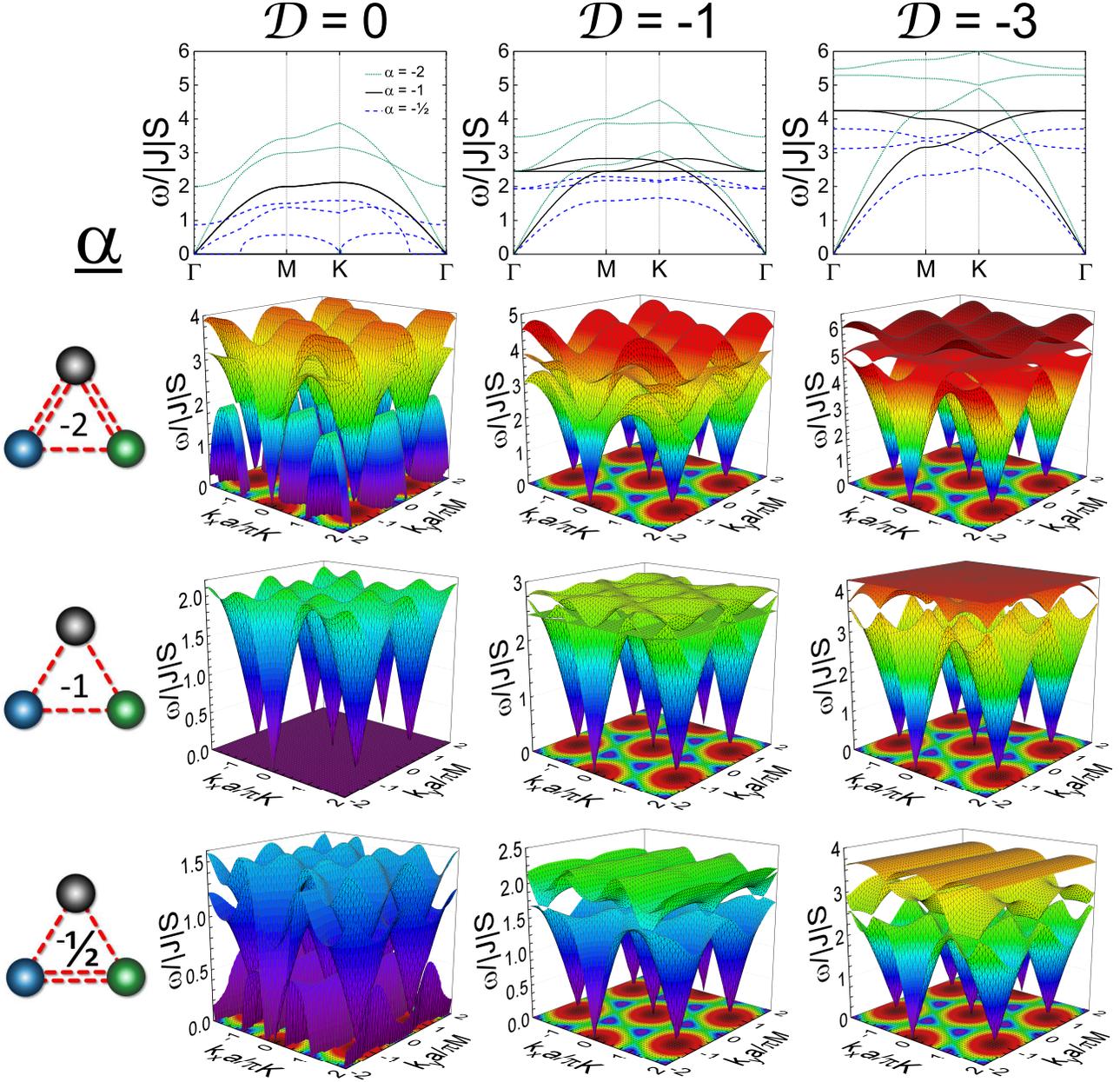}
    \caption{Isosceles distortions of spin waves for the 120$^\circ$ configuration. The first row shows the spin-wave energy values through the path in $k$-space that passes through symmetry points $\Gamma$, M, K, then back to $\Gamma$. The $\alpha$ value for each row of 3D graphs is given by the diagram in the left-hand column. Note that there are no stable spin waves for this configuration without anisotropy, aside from the equilateral ($\alpha = -1$) mode. This being the case, multiple magnitudes of anisotropy were included to better illustrate the stable possibilities for the 120$^{\circ}$ phase. Because no positive-$\alpha$ state could be stabilized with anisotropy lesser in magnitude than $3|J|$, only negative values were included here. A single color group encompasses all 3D graphs in this figure.}
    \label{Isosceles_Spinwaves_120}
\end{figure*}

In the isosceles case, the minimizing relationship becomes 
\begin{equation} \label{alpha_min_E_IN_Isosceles}
    \displaystyle \alpha = 
    \mathcal{J}\frac{\sin\Big(\frac{\pi}{3}+2g\Big)}{\cos\Big(\frac{\pi}{6}+g\Big)}.
\end{equation}

Equation (\ref{alpha_min_E_IN_Isosceles}), illustrated in Fig. \ref{AlphaVsQG}, shows that to minimize $E_0$, the absolute value of exchange interaction $J'$ may never exceed twice the value of $J$. It may also be observed that when $J$ is positive, the maximum value $J' = J$ requires the AfM configuration ($g$ = $\frac{\pi}{3}$) to minimize classical energy and the minimum value $J'$ = $-J$ requires the FM configuration ($g$ = $-\frac{2\pi}{3}$).

Examination of the equilateral case showed that the minima and maxima of $E_0/ | J |S^2$, regardless of the sign of $J$, are produced by the 120$^{\circ}$ and FM configurations, respectively. The 120$^{\circ}$ phase has the lowest energy of all equilateral structures in and out of the plane and is determined as the ground state. 

The AfM configuration features as a local minimum for the classical energy, yet proved too unstable for spin-wave examination in the equilateral case. It is analyzed, however, with further distortions.

The FM, AfM, and 120$^\circ$ phases will therefore be the initial subjects for in-plane analyses. In the plane, the ferromagnetic energy is simplified to
\begin{equation} \label{E_0_IN_FM}
    \displaystyle \frac{E_{0,FM}}{| J |S^2} = -\frac{2}{3}(\alpha + \beta + \mathcal{J}),
\end{equation} 
the 120$^{\circ}$ configuration energy to
\begin{equation} \label{E_0_IN_120}
    \displaystyle \frac{E_{0,120}}{| J |S^2} = \frac{1}{3}(\alpha + \beta + \mathcal{J}),
\end{equation} 
and the AfM energy to
\begin{equation} \label{E_0_AfM}
    \displaystyle \frac{E_{0,AfM}}{| J |S^2} = \frac{1}{3}(-\alpha - \beta + \mathcal{J}).
\end{equation}

Overall, the classical energies allow for the general understanding of where these few configurations are stable with respect to each other. It is important to note that other magnetic structures, especially canted non-collinear or $>$3 SL magnetic configurations could also exist. However, there are too many to meaningfully characterize all possibilities here. Therefore, it is essential to look at the spin dynamics to gain insight into the spin configurations' stability. If a system is a stable ground state according to the classical considerations but unstable from the standpoint of the spin dynamics, then this is an indication of the presence of a canted non-collinear state. Therefore, the next step is to evaluate these configurations for dynamic stability.

\section{Spin-wave Dynamics}

Exploring the spin-wave dynamics for the five most interesting spin configurations determined by the classical energy, we first generate solutions to the simplest case where all exchange interactions are equal, no magnetic field is applied, the spin angles are held constant, the physical distance between each atom remains static, and the only variation is anisotropy. Beyond this, we produce spin-wave solutions for the sublattice where exchange interactions are no longer equal, first exploring only one distortion (the isosceles case), then examining the case where no two interactions are equal, termed the ``scalene" case. Considering purely in-plane spin configurations, the latter two cases are further probed by varying the spin angles. The varied spin angles considered belong to the classical energy minimizing relationships discussed in the previous section. 

There are a few formatting rules applied throughout the figures in this next section. For every 3-D spin-wave graph, a color scale is applied which spans all colors from purple to red, where purple reflects the minimum $\omega/|J|S$ value 0 and red indicates the maximum $\omega/|J|S$ value between all graphs in the group of graphs to which that color scale is applied. For example, in Fig. \ref{Equilateral_Spinwaves}, the $120^\circ$ graphs (k) and (l) are grouped, and the color scale for both spans from purple at 0 to red at 4.5, where 4.5 is the highest $\omega/|J|S$ value between both graphs. The color grouping used between graphs is specified in each figure caption. 

To contextualize the spin waves' form with respect to the reciprocal lattice, all 3D spin-wave graphs include a projection of the equilateral FM configuration's central energy level (pictured in Fig. \ref{Equilateral_Spinwaves}(a)) onto the $k$-plane. The projection is a heat map whose colors are scaled relative to that single energy level's minima and maxima, as in Fig. \ref{brillouin}(b). 

The 2D spin-wave graphs in the isosceles figures include only three $\alpha$ values, as opposed to the six considered in the 3D graphs, because the inclusion of all six $\alpha$ values created massive overlap in the lines defining the energies, rendering the graphs unreadable. Any destabilizing $\alpha$ values were thus excluded from these representations to offer greater clarity.

All spin-wave figures after the equilateral figure include small triangular diagrams. These are included to visualize the proportions of the exchange interaction strengths between atoms. Red lines indicate negative valued interactions and black positive. The atoms are colored to reflect the spin configurations as in Fig. \ref{structure}, save those in the isosceles graphs: as the FM and AfM configurations share triangle diagrams, the diagrams were colored to reflect the AfM mode.

\begin{figure*}[!]
    \includegraphics[width=\textwidth]{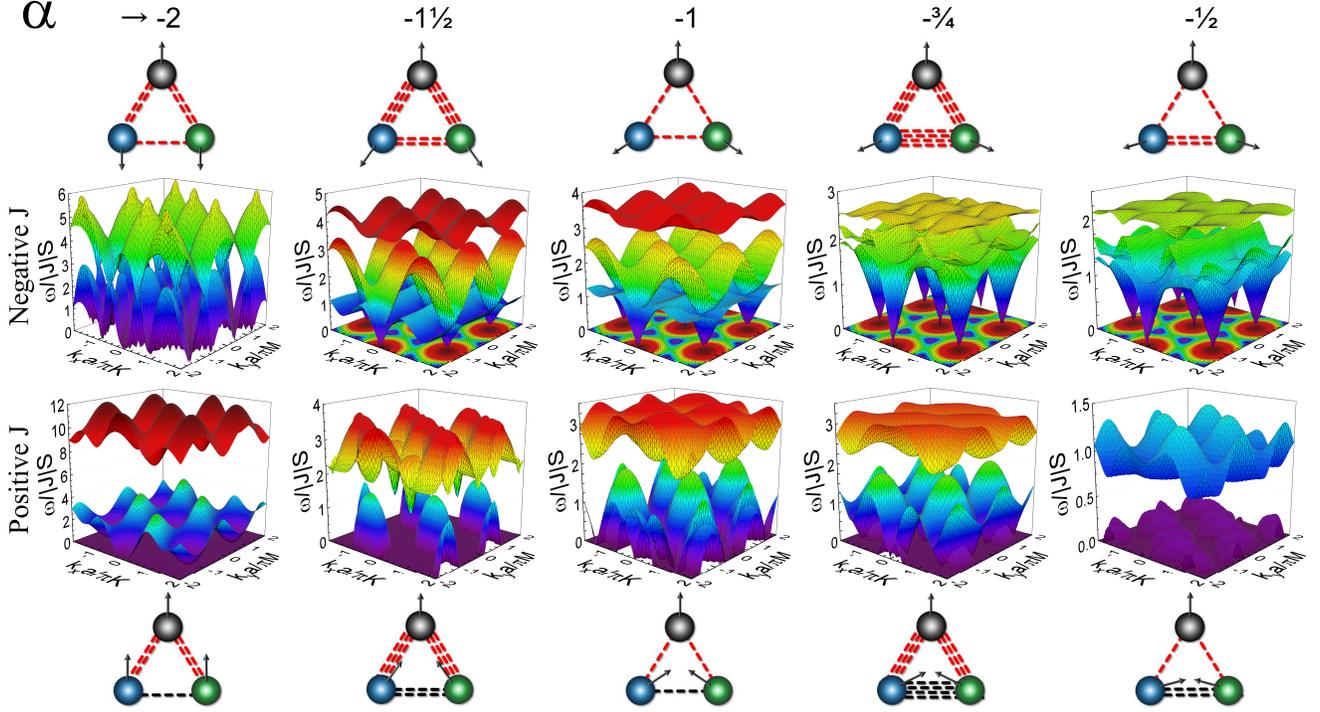}
    \caption{Isosceles distortions of in-plane spin waves according to the relationship defined in Eq. \ref{alpha_min_E_IN_Isosceles}. Aside from the two graphs whose $\alpha$ value approaches 2, all spin waves pictured have anisotropy $\mathcal{D}$ = 1. Only negative $\alpha$ values are depicted because when both the $\alpha$ and $J$ signs are reversed for any given combination, they produce the same spin waves, with the exception of the addition of anisotropy. Additionally, the spin rotating angles are provided in Table \ref{isosceles_table}, and two groups are defined for clarity in color scaling, where those without anisotropy are grouped together and those with are grouped separately.}
    \label{Emin_Isosceles_Spinwaves}
\end{figure*}

\subsection{Out-of-plane Configurations}

\subsubsection{Ferromagnetic Phase}

Since all spins in the FM phase point in the same direction, it is expected this phase is stable when each of the exchange interactions between these aligned sites ($\mathcal{J}$, $\alpha$, and $\beta$) are positive. We examine the effects of interaction competition on the FM phase to understand when it becomes unstable from a spin-wave standpoint.

Figure \ref{Equilateral_Spinwaves}(b) and (c) show how the equilateral ($\alpha$ = $\beta$ = $\mathcal{J}$ = +1) out-of-plane FM spin waves evolve with anisotropy. The anisotropy present in Fig. \ref{Equilateral_Spinwaves}(c) is equal to the exchange interaction ($\mathcal{D}$ = +1). It is observed that, without anisotropy, the spin waves are identical to the FM configuration pointing in the plane Fig. \ref{Equilateral_Spinwaves}(h). As expected, easy-axis anisotropy present in the out-of-plane configuration simply adds uniformly across the $k$-plane to all energy states. The spin-wave energies $\omega_i$ are represented analytically as

\begin{widetext}
\begin{equation}
\begin{array}{ll}
    \omega_0/|J|S = & 2\mathcal{D}+6\mathcal{J} \\
    \omega_{\pm}/|J|S = & 2\mathcal{D}+3\mathcal{J}\pm\sqrt{8\mathcal{J}^2\cos(k_x)^2\cos(\frac{k_x}{2}-\frac{\sqrt{3}k_y}{2})^2+8\mathcal{J}^2\cos(\frac{k_x}{2}\pm \frac{\sqrt{3}k_y}{2})\cos(k_x)\sin(k_x)\sin(\frac{k_x}{2}-\frac{\sqrt{3}k_y}{2})+\mathcal{J}^2}. \\
\end{array}
\end{equation}
\end{widetext}

Moving to the first distortion of exchange interactions, the spin-wave dynamics of various isosceles states ($\alpha$ = $\beta$) are explored for the FM configuration in Fig. \ref{Isosceles_Spinwaves_Out}, which illustrates the transformation of the spin waves as the value $\alpha$ changes. With an $\alpha$ value of 2, the interaction between atoms B and C has half the magnitude of the other two. This distortion is reflected in the dispersion shown in the $k_y$ direction when inspecting the spin-wave diagram. A similar, perpendicular phenomenon can be seen in the $\alpha = \frac{1}{2}$ case. The inclusion of anisotropy energy, predictably, has a similar additive effect to that observed in the equilateral spin waves. As the spins are uni-directional in a ferromagnet, it is expected that the exchange interactions are all positive. This instability is confirmed visually by inspecting the negative $\alpha$-valued graphs, except for the $\alpha$ = -2 spin waves, which seem to become stable due to a symmetry effect in the interactions. Larger negative $\alpha$ values produce equally stable-looking waves. It becomes apparent, however, that this state is not truly stable when considering its behavior with anisotropy. Notice that the graph in the second column, which reflects $\mathcal{D} = +1$, has lower overall energy than its no-anisotropy counterpart, which is indicative of an unstable system. This is in agreement with the classical energy, which indicates that this state is only metastable. A similar effect in the AfM configuration is realized in the next analysis. While calculable, the analytical representation of the isosceles spin waves is too large to be presented here and cannot be used to provide insight. 

Figure \ref{Scalene_Spinwaves_FM} shows the scalene spin waves with all permutations of a 1-2-3 proportionality between $\alpha$, $\beta$, and $J$, given a static FM spin configuration. It is observed that, as in the isosceles case, both the in-plane and out-of-plane FM configurations produce identical spin waves throughout the variation of exchange interaction, so long as anisotropy is not considered. However, their changes with respect to anisotropy differ. For simplicity, anisotropy was excluded from the scalene discussion. These visualizations illustrate the dispersion that occurs along the direction of the most robust exchange interaction.

\subsubsection{Antiferrimagnetic Phase}

In this configuration, sites B and C in Fig. \ref{brillouin}(a) are aligned and site A points in the opposite direction. It is thus expected that this configuration will be stabilized by a positive exchange interaction $J$ and negative interactions $J'$ and $J''$. However, with frustration in this system, the boundaries of stability are of interest, as with the FM phase.

The out-of-plane, equilateral AfM energies are unstable without anisotropy. The anisotropies $\mathcal{D}$ present in Fig. \ref{Equilateral_Spinwaves} (e) and (f) are valued at +1 and +2, respectively. As discussed prior, it is expected that the AfM configuration contains one positive exchange interaction (between the two same-spin atoms) and two negative interactions. However, as all exchange interactions are strictly equal in the equilateral case, all exchange interactions were assigned a negative value to reflect the net negativity of the three interactions together. Therefore, this state has $\mathcal{J} = \alpha = \beta = -1$. The analytical representation of these energies is too large for inclusion.

Figure \ref{Isosceles_Spinwaves_Out} includes the isosceles ($\alpha$ = $\beta$) out-of-plane AfM spin waves. The expected stabilizing state is $\alpha$ = -1, while $J$ is positive, as the exchange interactions $\alpha$ and $\beta$ are equal and belong to the opposite-spin atom pairs. The spin waves at this $\alpha$ value reflect this expectation in their visually apparent stability. Interestingly, it is shown that various $\alpha$ values that stabilize this otherwise unstable configuration. Similar to the FM phase, the AfM system has an unexpected metastable state where all three exchange interactions are positive with an $\alpha$ value of +2. Again, we see an overall lowering of energy when anisotropy is included, revealing the volatile nature of this arrangement. 

The scalene spin waves with all permutations of a 1-2-3 proportionality between $\alpha$, $\beta$, and $J$, for an AfM spin configuration, are shown in Fig. \ref{Scalene_Spinwaves_AfM}. As in the isosceles case, the in-plane and out-of-plane configurations, with no anisotropy, produce identical spin waves throughout the exchange interaction variation. For the sake of simplicity, anisotropy-inclusive spin waves are excluded from the narrative. As with the isosceles case, dispersion in the spin waves occurs along the direction of the strongest exchange interaction. 

\begin{table}
\begin{tabular}{|p{1cm}|p{1cm}|p{1cm}||p{1cm}|p{1cm}|p{1cm}|}
 \hline
 \multicolumn{6}{|c|}{Isosceles spins} \\
 \hline
 $\mathcal{J}$&$\alpha$&$\beta$&$\phi_A (^{\circ}$)&$\phi_B(^{\circ}$)&$\phi_C(^{\circ}$)\\
 \hline
 -1& -2   & -2   & 90 & 270.0 & 270.0\\
 -1& -3/2 & -3/2 & 90 & 228.6 & 311.4\\
 -1& -1   & -1   & 90 & 210.0 & 330.0\\
 -1& -3/4 & -3/4 & 90 & 202.0 & 338.0\\
 -1& -1/2 & -1/2 & 90 & 194.5 & 345.5\\
 1&  -2   & -2   & 90 & 90.00 & 90.00\\
 1&  -3/2 & -3/2 & 90 & 131.4 & 48.60\\
 1&  -1   & -1   & 90 & 150.0 & 30.00\\
 1&  -3/4 & -3/4 & 90 & 158.0 & 22.00\\
 1&  -1/2 & -1/2 & 90 & 165.5 & 14.50\\
 \hline
\end{tabular}
\caption{\label{tab:table-name}The spin angles calculated for $\alpha$ values by the relationship in Eq. \ref{alpha_min_E_IN_Isosceles}}
\label{isosceles_table}
\end{table}

\subsection{In-plane Configurations}

\subsubsection{Ferromagnetic Phase}

Figure \ref{Equilateral_Spinwaves} (h) and (i) show the evolution of the equilateral in-plane FM spin waves with the presence of anisotropy.  Easy-plane anisotropy added to the in-plane configuration stretches the spin waves as interactions strain the easy orientation. For this configuration, the spin-wave energies can be represented analytically and shown to be
\begin{widetext}
\begin{equation}
\begin{array}{ll}
    \omega_0/|J|S = & 2\sqrt{9\mathcal{J}^2-3\mathcal{D}\mathcal{J}} \\
    \omega_{\pm}/|J|S = & \biggl(8\mathcal{J}^2\sin(k_x)\cos(\frac{k_x}{2}-\frac{\sqrt{3}k_y}{2})\sin(\frac{k_x}{2}-\frac{\sqrt{3}k_y}{2})\cos(k_x)+8\mathcal{J}^2(\cos(\frac{k_x}{2}-\frac{\sqrt{3}k_y}{2}))^2(\cos(k_x))^2-6\mathcal{D}\mathcal{J}+10\mathcal{J}^2 \\
    &\pm 2\mathcal{J}\sqrt{8(\mathcal{D}-3\mathcal{J})^2(\sin(k_x)\cos(\frac{k_x}{2}-\frac{\sqrt{3}k_y}{2})\sin(\frac{k_x}{2}-\frac{\sqrt{3}k_y}{2})\cos(k_x)+\cos(\frac{k_x}{2}-\frac{\sqrt{3}k_y}{2})^2(\cos(k_x))^2+1)}\biggr)^\frac{1}{2}. \\
\end{array}
\end{equation}
\end{widetext}

Various isosceles FM spin waves are illustrated in Fig. \ref{Isosceles_Spinwaves_In}. It is shown that, as in the equilateral case, the spin waves are identical to the out-of-plane configuration where anisotropy is not considered. Otherwise, the inclusion of anisotropy energy has the same stretching effect as that in the equilateral spin waves. Like the out-of-plane FM configuration, the analytical representation of the energies that describe an isosceles in-plane FM state were too large to include here.

As no anisotropy was considered for the scalene cases, Fig. \ref{Scalene_Spinwaves_FM} represents both in- and out-of-plane configurations and is not discussed redundantly.

\subsubsection{Antiferrimagnetic Phase}

Even with the presence of anisotropy five times the magnitude of the exchange interaction, the equilateral spin waves for the in-plane AfM configuration were unstable and therefore excluded from figure \ref{Equilateral_Spinwaves}. 

Figure \ref{Isosceles_Spinwaves_In} illustrates isosceles in-plane AfM spin waves. As in the out-of-plane case, there are various $\alpha$ values which stabilize this unstable configuration. Without anisotropy, these spin waves are identical to the out-of-plane spin waves. Yet, when present, a stretching of energy is observed as a result of anisotropy.

Scalene spinwaves for the in-plane AfM configuration with $\mathcal{D} = 0$ are shown in figure \ref{Scalene_Spinwaves_AfM} and are identical to those of their out-of-pane counterparts.


\begin{figure*}[!]
    \includegraphics[width=\textwidth]{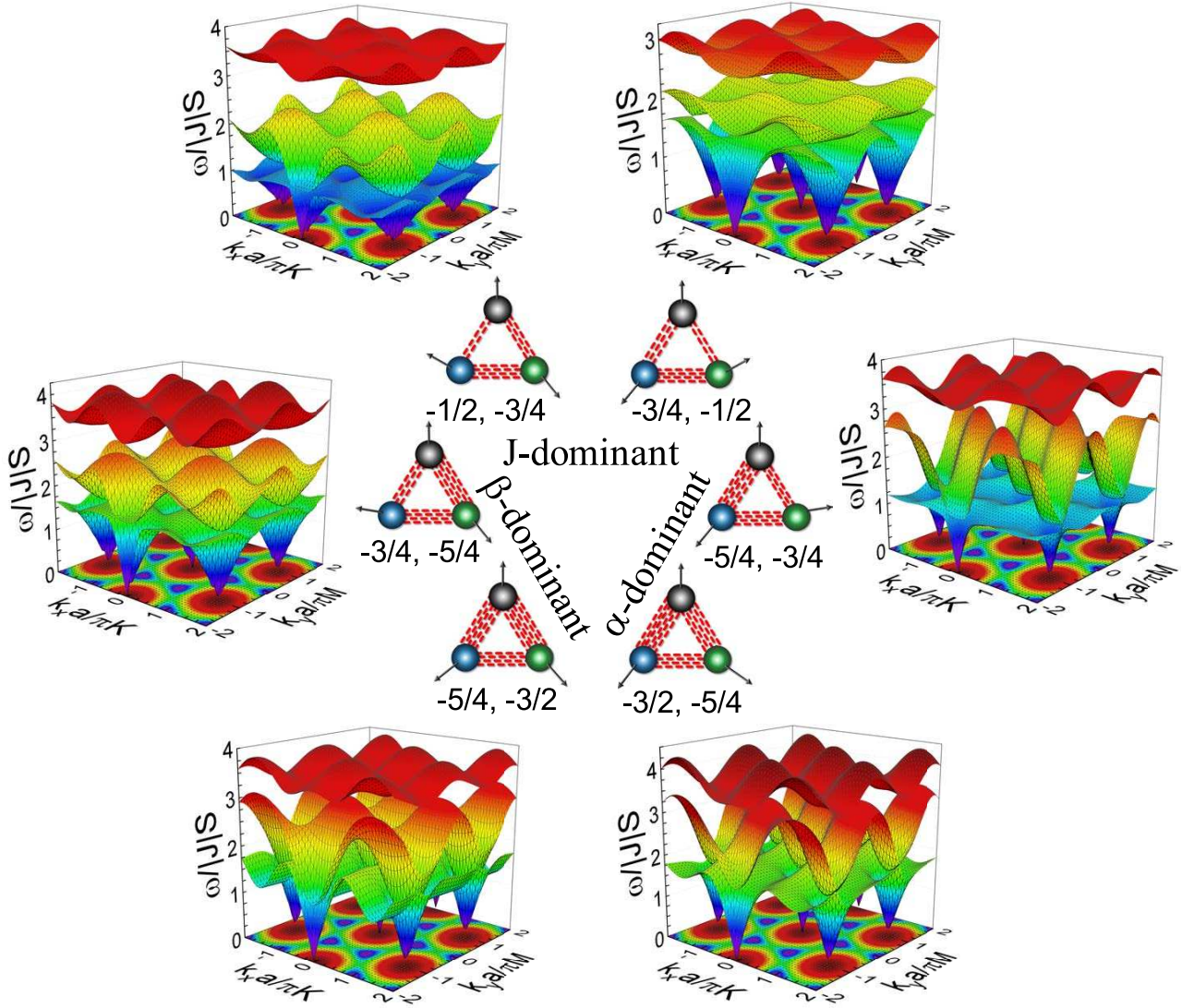}
    \caption{Scalene distortions of in-plane spin waves according to the relationships in Eq. \ref{alpha_beta_min_E_IN}. All spin waves shown have anisotropy $\mathcal{D}$ = 1. The $\alpha$ and $\beta$ values for each graph are presented within the triangle diagrams, respectively. Additionally, the spin rotating angles are provided in Table \ref{scalene_table}. A single color group encompasses all graphs.}
    \label{Emin_Scalene_Spinwaves}
\end{figure*}

\subsubsection{120$^{\circ}$ Phase}

The ground state of the Kagome lattice behaves peculiarly in contrast to the behaviors of the FM and AfM configurations. The no-anisotropy spin waves in Fig. \ref{Equilateral_Spinwaves} (k) show a degeneracy of energy levels for the 120$^{\circ}$ phase. This degeneracy is lost, however, as anisotropy increases. As anisotropy energy increases to outweigh the exchange interactions, Dirac nodes form in the system. While lower values of anisotropy maintained stable spin waves, the relatively large amount of $\mathcal{D}$ = -3 is included for the purpose of illustrating this phenomenon. Lesser anisotropies are included in the 2-D spin waves graph (j) to convey its interesting path to this point. Here, the analytical spin-wave energies can be shown as

\begin{widetext}
\begin{equation}
\begin{array}{ll}
    \omega_0/|J|S = & \sqrt{6\mathcal{J}\mathcal{D}} \\
    \omega_{\pm}/|J|S = & \biggl(-4\mathcal{J}^2\cos(\frac{k_x}{2}-\frac{\sqrt{3}k_y}{2})^2\cos(k_x)^2-4\mathcal{J}^2\cos(\frac{k_x}{2}-\frac{\sqrt{3}k_y}{2})\sin(\frac{k_x}{2}-\frac{\sqrt{3}k_y}{2})\cos(k_x)\sin(k_x)+3\mathcal{D}\mathcal{J}+4\mathcal{J}^2 \\
    & \pm \mathcal{J}\sqrt{(8\cos(\frac{k_x}{2}-\frac{\sqrt{3}k_y}{2})^2\cos(k_x)^2+8\cos(\frac{k_x}{2}-\frac{\sqrt{3}k_y}{2})\sin(\frac{k_x}{2}-\frac{\sqrt{3}k_y}{2})\cos(k_x)\sin(k_x)+1)\mathcal{D}^2}\biggr)^\frac{1}{2} . \\
\end{array}
\end{equation}
\end{widetext}

The distortion of exchange energies did not provide any additional stability to the 120$^{\circ}$ configuration. In fact, any distortion destabilized the system and required anisotropy on the order of the exchange interaction for stability. Even anisotropy, however, could not stabilize any state whose $\alpha$ value had a sign opposite to that of $J$. This instability is predicted by the state being a particular case of energy minimization related to the $\alpha$ and $\beta$ values.

Because isosceles distorting exchange interactions for the 120$^{\circ}$ phase offered no stability, scalene spin wave analyses for this configuration were omitted entirely.

\begin{table}
\begin{tabular}{|p{1cm}|p{1cm}|p{1cm}||p{1cm}|p{1cm}|p{1cm}|}
 \hline
 \multicolumn{6}{|c|}{Scalene spins} \\
 \hline
 $\mathcal{J}$&$\alpha$&$\beta$&$\phi_A (^{\circ}$)&$\phi_B(^{\circ}$)&$\phi_C(^{\circ}$)\\
 \hline
-1& -5/4 & -3/2 & 90 & 217.1 & 311.6\\
-1& -3/4 & -5/4 & 90 & 175.1 & 306.7\\
-1& -1/2 & -3/4 & 90 & 152.7 & 306.3\\
-1& -3/4 & -1/2 & 90 & 233.7 & 27.27\\
-1& -5/4 & -3/4 & 90 & 233.3 & 4.940\\
-1& -3/2 & -5/4 & 90 & 228.4 & 322.9\\
 \hline
\end{tabular}
\caption{\label{tab:table-name}The spin angles calculated for various combinations of $\alpha$ and $\beta$ determined by the relationships in Eq. \ref{alpha_beta_min_E_IN}}
\label{scalene_table}
\end{table}

\subsection{Spin Angle Distortions}

The 120$^\circ$ phase showed to be quite a unique configuration. While it is the ground state when all exchange interactions are equal, any distortion of the exchange parameters destabilize the phase, unlike the other configurations which could maintain stability through some distortion. In light of this, we looked for stability in new states whose spin angles and exchange interactions together minimize the classical energy.

Before considering all distorting parameters, we investigate the isosceles ($\alpha$ = $\beta$) states under the assumption of the simplest case, where rotation angles $g$ and $h$ are also equivalent, as described in Eq. \ref{alpha_min_E_IN_Isosceles}. It can be inferred from this minimizing relationship, illustrated in Fig. \ref{AlphaVsQG}, that there are two angle $g$ solutions for any given $\alpha$. However, these two values produce equivalent spin waves as the two $g$ values produce the same two values for spin angles for atoms B and C, but the angles assigned to each site are reversed. Spinwaves for some interesting $\alpha$ values are illustrated in Fig. \ref{Emin_Isosceles_Spinwaves}. As nearly every result was unstable, an anisotropy of $\mathcal{D} = -1$ was applied to all states except those two which have an $\alpha$ value approaching 2. These do not include anisotropy. 

Examining further distortion, figure \ref{Emin_Scalene_Spinwaves} shows scalene spin waves with various values of $\alpha$ and $\beta$, given a negative $J$, and with spin angles defined by the classical energy minimizing relationships in Eq. \ref{alpha_beta_min_E_IN}. As the scalene spin waves were also unstable without anisotropy, the solutions pictured reflect a value $\mathcal{D} = -1$. These visualizations illustrate the dispersion that occurs along the direction of the strongest exchange interaction. The fact that these polar spin angle solutions which minimize the classical energy rely on anisotropy for stability may indicate that the ground state configurations for these distorted exchange interaction states are canted out of the lattice plane to some degree.

Overall, these results indicate that purely in-plane magnetic configurations which deviate from the FM, AfM, and 120$^\circ$ states require anisotropy to stabilize within the Kagome lattice. We expect that there will be a reduction in the required anisotropy as the spin angles are allowed to cant in any out-of-plane direction (not necessarily along the $z$-axis) and form distinct non-collinear phases. Further study into the realm of these out-of-plane canted phases needs to be done. However, the results here are important for understanding the evolution of the magnetic Kagome lattice's spin waves with variable exchange interactions.

\section{Discussion and Conclusion}

The quest for an understanding of quantum spin states, especially the quantum spin liquid, has led to a wealth of experimental realizations and studies on the structural, magnetic, and thermodynamics of Kagome systems\cite{han:12:nature,helt:07:PRL,hirs:15:PRL,dun:16:PRL,kass:17:PRB,zork:19:PRB} as well as distorted Kagome systems\cite{wei:10:PRB,wulf:12:JPCM,mata:19:PRB,mata:10:NatPhys}. While many studies focus on the interpretation of either one material or even one magnetic configuration, the ability to discern the various magnetic interactions with and without distortions has been a challenge that leads most to using numerical approaches in modeling experimental data. However, examining these systems within an analytical limit for the symmetric systems and evolving out into the distorted systems allows for one to gain a deeper understanding of the effects of the interactions and how they distort the magnetic systems.

In this paper, we examine the effects of variable first-order magnetic distortions on the spin-wave dynamics of the Kagome lattice. Using an isotropic spin-spin exchange Hamiltonian with $z$-axis anisotropy, we determine the phase diagrams for various out-of-plane and in-plane spin configurations and then examine how the spin waves are affected by exchange interaction distortions. 

By analyzing the spin waves with static spin angles and varying exchange interactions, we gain novel insights on the effect this distortion alone has on the magnetic identity of this sublattice. A dispersive effect on the spin waves in accordance with exchange interaction proportionality is illustrated in the isosceles and scalene cases for all five of the most straightforward configurations, and some unexpectedly metastable states fell out of unlikely exchange interaction states for the FM and AfM cases. 

Although it is the ground state classically, the special-case nature of the 120$^\circ$ phase is underlined by its inability to retain stability as its exchange interactions are distorted. This led to our employing the energy-minimizing relationships for spin angles in seeking stable spin waves for these changing $\alpha$ values, which pointed to even further spin distortions for stability.  The next natural step in this vein would be to numerically minimize the classical energy with full freedom of spin angles, allowing for continuity in both the polar and azimuthal angles. Limiting to purely in-plane or out-of-plane angles offered analytical insights that are invaluable in characterizing magnetic relationships. However, to more deeply describe the most natural behaviors of atoms in this lattice, the allowance of canted configurations is a necessary consideration.

Overall, this paper aims to provide insight into the spin dynamics of the Kagome lattice to help in the characterization of its non-collinear phases. Therefore, we show how the distortion of the exchange parameters and spin angles affect the dynamics, which provides useful information for the characterization of material systems, especially when investigating these phases using inelastic neutron scattering. 

\section*{Acknowledgements}

A.A.C. and J.T.H acknowledge support from the Institute for Materials Science at Los Alamos National Laboratory. The work at Los Alamos National Laboratory was carried out under the auspices of the U.S. DOE and NNSA under Contract No. DEAC52-06NA25396 and supported by U.S. DOE  (A.S.).

\bibliography{kagome.bib}

\begin{thebibliography}{50}%
\makeatletter
\providecommand \@ifxundefined [1]{%
 \@ifx{#1\undefined}
}%
\providecommand \@ifnum [1]{%
 \ifnum #1\expandafter \@firstoftwo
 \else \expandafter \@secondoftwo
 \fi
}%
\providecommand \@ifx [1]{%
 \ifx #1\expandafter \@firstoftwo
 \else \expandafter \@secondoftwo
 \fi
}%
\providecommand \natexlab [1]{#1}%
\providecommand \enquote  [1]{``#1''}%
\providecommand \bibnamefont  [1]{#1}%
\providecommand \bibfnamefont [1]{#1}%
\providecommand \citenamefont [1]{#1}%
\providecommand \href@noop [0]{\@secondoftwo}%
\providecommand \href [0]{\begingroup \@sanitize@url \@href}%
\providecommand \@href[1]{\@@startlink{#1}\@@href}%
\providecommand \@@href[1]{\endgroup#1\@@endlink}%
\providecommand \@sanitize@url [0]{\catcode `\\12\catcode `\$12\catcode
  `\&12\catcode `\#12\catcode `\^12\catcode `\_12\catcode `\%12\relax}%
\providecommand \@@startlink[1]{}%
\providecommand \@@endlink[0]{}%
\providecommand \url  [0]{\begingroup\@sanitize@url \@url }%
\providecommand \@url [1]{\endgroup\@href {#1}{\urlprefix }}%
\providecommand \urlprefix  [0]{URL }%
\providecommand \Eprint [0]{\href }%
\providecommand \doibase [0]{https://doi.org/}%
\providecommand \selectlanguage [0]{\@gobble}%
\providecommand \bibinfo  [0]{\@secondoftwo}%
\providecommand \bibfield  [0]{\@secondoftwo}%
\providecommand \translation [1]{[#1]}%
\providecommand \BibitemOpen [0]{}%
\providecommand \bibitemStop [0]{}%
\providecommand \bibitemNoStop [0]{.\EOS\space}%
\providecommand \EOS [0]{\spacefactor3000\relax}%
\providecommand \BibitemShut  [1]{\csname bibitem#1\endcsname}%
\let\auto@bib@innerbib\@empty
\bibitem [{\citenamefont {Kitaev}(2006)}]{kita:06:AoP}%
  \BibitemOpen
  \bibfield  {author} {\bibinfo {author} {\bibfnamefont {A.}~\bibnamefont
  {Kitaev}},\ }\bibfield  {title} {\bibinfo {title} {Anyons in an exactly
  solved model and beyond},\ }\href
  {https://doi.org/https://doi.org/10.1016/j.aop.2005.10.005} {\bibfield
  {journal} {\bibinfo  {journal} {Annals of Physics}\ }\textbf {\bibinfo
  {volume} {321}},\ \bibinfo {pages} {2 } (\bibinfo {year} {2006})},\ \bibinfo
  {note} {january Special Issue}\BibitemShut {NoStop}%
\bibitem [{\citenamefont {Knolle}\ and\ \citenamefont
  {Moessner}(2019)}]{knol:19:ARoCMP}%
  \BibitemOpen
  \bibfield  {author} {\bibinfo {author} {\bibfnamefont {J.}~\bibnamefont
  {Knolle}}\ and\ \bibinfo {author} {\bibfnamefont {R.}~\bibnamefont
  {Moessner}},\ }\bibfield  {title} {\bibinfo {title} {A field guide to spin
  liquids},\ }\href {https://doi.org/10.1146/annurev-conmatphys-031218-013401}
  {\bibfield  {journal} {\bibinfo  {journal} {Annual Review of Condensed Matter
  Physics}\ }\textbf {\bibinfo {volume} {10}},\ \bibinfo {pages} {451}
  (\bibinfo {year} {2019})}\BibitemShut {NoStop}%
\bibitem [{\citenamefont {Chen}\ and\ \citenamefont
  {Balents}(2008)}]{gang:08:PRB}%
  \BibitemOpen
  \bibfield  {author} {\bibinfo {author} {\bibfnamefont {G.}~\bibnamefont
  {Chen}}\ and\ \bibinfo {author} {\bibfnamefont {L.}~\bibnamefont {Balents}},\
  }\bibfield  {title} {\bibinfo {title} {{Spin-orbit effects in
  ${\text{Na}}_{4}{\text{Ir}}_{3}{\text{O}}_{8}$: A hyper-kagome lattice
  antiferromagnet}},\ }\href {https://doi.org/10.1103/PhysRevB.78.094403}
  {\bibfield  {journal} {\bibinfo  {journal} {Phys. Rev. B}\ }\textbf {\bibinfo
  {volume} {78}},\ \bibinfo {pages} {094403} (\bibinfo {year}
  {2008})}\BibitemShut {NoStop}%
\bibitem [{\citenamefont {Chisnell}\ \emph {et~al.}(2015)\citenamefont
  {Chisnell}, \citenamefont {Helton}, \citenamefont {Freedman}, \citenamefont
  {Singh}, \citenamefont {Bewley}, \citenamefont {Nocera},\ and\ \citenamefont
  {Lee}}]{chis:15:PRL}%
  \BibitemOpen
  \bibfield  {author} {\bibinfo {author} {\bibfnamefont {R.}~\bibnamefont
  {Chisnell}}, \bibinfo {author} {\bibfnamefont {J.~S.}\ \bibnamefont
  {Helton}}, \bibinfo {author} {\bibfnamefont {D.~E.}\ \bibnamefont
  {Freedman}}, \bibinfo {author} {\bibfnamefont {D.~K.}\ \bibnamefont {Singh}},
  \bibinfo {author} {\bibfnamefont {R.~I.}\ \bibnamefont {Bewley}}, \bibinfo
  {author} {\bibfnamefont {D.~G.}\ \bibnamefont {Nocera}},\ and\ \bibinfo
  {author} {\bibfnamefont {Y.~S.}\ \bibnamefont {Lee}},\ }\bibfield  {title}
  {\bibinfo {title} {{Topological Magnon Bands in a Kagome Lattice
  Ferromagnet}},\ }\href {https://doi.org/10.1103/PhysRevLett.115.147201}
  {\bibfield  {journal} {\bibinfo  {journal} {Phys. Rev. Lett.}\ }\textbf
  {\bibinfo {volume} {115}},\ \bibinfo {pages} {147201} (\bibinfo {year}
  {2015})}\BibitemShut {NoStop}%
\bibitem [{\citenamefont {Haraldsen}\ \emph {et~al.}(2009)\citenamefont
  {Haraldsen}, \citenamefont {Swanson}, \citenamefont {Alvarez},\ and\
  \citenamefont {Fishman}}]{hara:09:PRL}%
  \BibitemOpen
  \bibfield  {author} {\bibinfo {author} {\bibfnamefont {J.~T.}\ \bibnamefont
  {Haraldsen}}, \bibinfo {author} {\bibfnamefont {M.}~\bibnamefont {Swanson}},
  \bibinfo {author} {\bibfnamefont {G.}~\bibnamefont {Alvarez}},\ and\ \bibinfo
  {author} {\bibfnamefont {R.~S.}\ \bibnamefont {Fishman}},\ }\bibfield
  {title} {\bibinfo {title} {Spin-wave instabilities and noncollinear magnetic
  phases of a geometrically frustrated triangular-lattice antiferromagnet},\
  }\href {https://doi.org/10.1103/PhysRevLett.102.237204} {\bibfield  {journal}
  {\bibinfo  {journal} {Phys. Rev. Lett.}\ }\textbf {\bibinfo {volume} {102}},\
  \bibinfo {pages} {237204} (\bibinfo {year} {2009})}\BibitemShut {NoStop}%
\bibitem [{\citenamefont {Haraldsen}\ and\ \citenamefont
  {Fishman}(2009)}]{hara:09:JPCM}%
  \BibitemOpen
  \bibfield  {author} {\bibinfo {author} {\bibfnamefont {J.~T.}\ \bibnamefont
  {Haraldsen}}\ and\ \bibinfo {author} {\bibfnamefont {R.~S.}\ \bibnamefont
  {Fishman}},\ }\bibfield  {title} {\bibinfo {title} {{Spin rotation technique
  for non-collinear magnetic systems: application to the generalized Villain
  model}},\ }\href {https://doi.org/10.1088/0953-8984/21/21/216001} {\bibfield
  {journal} {\bibinfo  {journal} {Journal of Physics: Condensed Matter}\
  }\textbf {\bibinfo {volume} {21}},\ \bibinfo {pages} {216001} (\bibinfo
  {year} {2009})}\BibitemShut {NoStop}%
\bibitem [{\citenamefont {Mukherjee}\ \emph {et~al.}(2015)\citenamefont
  {Mukherjee}, \citenamefont {Spracklen}, \citenamefont {Choudhury},
  \citenamefont {Goldman}, \citenamefont {\"Ohberg}, \citenamefont
  {Andersson},\ and\ \citenamefont {Thomson}}]{mukh:15:PRL}%
  \BibitemOpen
  \bibfield  {author} {\bibinfo {author} {\bibfnamefont {S.}~\bibnamefont
  {Mukherjee}}, \bibinfo {author} {\bibfnamefont {A.}~\bibnamefont
  {Spracklen}}, \bibinfo {author} {\bibfnamefont {D.}~\bibnamefont
  {Choudhury}}, \bibinfo {author} {\bibfnamefont {N.}~\bibnamefont {Goldman}},
  \bibinfo {author} {\bibfnamefont {P.}~\bibnamefont {\"Ohberg}}, \bibinfo
  {author} {\bibfnamefont {E.}~\bibnamefont {Andersson}},\ and\ \bibinfo
  {author} {\bibfnamefont {R.~R.}\ \bibnamefont {Thomson}},\ }\bibfield
  {title} {\bibinfo {title} {{Observation of a Localized Flat-Band State in a
  Photonic Lieb Lattice}},\ }\href
  {https://doi.org/10.1103/PhysRevLett.114.245504} {\bibfield  {journal}
  {\bibinfo  {journal} {Phys. Rev. Lett.}\ }\textbf {\bibinfo {volume} {114}},\
  \bibinfo {pages} {245504} (\bibinfo {year} {2015})}\BibitemShut {NoStop}%
\bibitem [{\citenamefont {Yang}\ \emph {et~al.}(2020)\citenamefont {Yang},
  \citenamefont {Perkins}, \citenamefont {Ko\ifmmode~\mbox{\c{c}}\else
  \c{c}\fi{}}, \citenamefont {Lin},\ and\ \citenamefont
  {Rousochatzakis}}]{yang:20:PRR}%
  \BibitemOpen
  \bibfield  {author} {\bibinfo {author} {\bibfnamefont {Y.}~\bibnamefont
  {Yang}}, \bibinfo {author} {\bibfnamefont {N.~B.}\ \bibnamefont {Perkins}},
  \bibinfo {author} {\bibfnamefont {F.}~\bibnamefont
  {Ko\ifmmode~\mbox{\c{c}}\else \c{c}\fi{}}}, \bibinfo {author} {\bibfnamefont
  {C.-H.}\ \bibnamefont {Lin}},\ and\ \bibinfo {author} {\bibfnamefont
  {I.}~\bibnamefont {Rousochatzakis}},\ }\bibfield  {title} {\bibinfo {title}
  {{Quantum-classical crossover in the spin-$\frac{1}{2}$ Heisenberg-Kitaev
  kagome magnet}},\ }\href {https://doi.org/10.1103/PhysRevResearch.2.033217}
  {\bibfield  {journal} {\bibinfo  {journal} {Phys. Rev. Research}\ }\textbf
  {\bibinfo {volume} {2}},\ \bibinfo {pages} {033217} (\bibinfo {year}
  {2020})}\BibitemShut {NoStop}%
\bibitem [{\citenamefont {Zhou}\ \emph {et~al.}(2017)\citenamefont {Zhou},
  \citenamefont {Kanoda},\ and\ \citenamefont {Ng}}]{zhou:17:RMP}%
  \BibitemOpen
  \bibfield  {author} {\bibinfo {author} {\bibfnamefont {Y.}~\bibnamefont
  {Zhou}}, \bibinfo {author} {\bibfnamefont {K.}~\bibnamefont {Kanoda}},\ and\
  \bibinfo {author} {\bibfnamefont {T.-K.}\ \bibnamefont {Ng}},\ }\bibfield
  {title} {\bibinfo {title} {Quantum spin liquid states},\ }\href
  {https://doi.org/10.1103/RevModPhys.89.025003} {\bibfield  {journal}
  {\bibinfo  {journal} {Rev. Mod. Phys.}\ }\textbf {\bibinfo {volume} {89}},\
  \bibinfo {pages} {025003} (\bibinfo {year} {2017})}\BibitemShut {NoStop}%
\bibitem [{\citenamefont {Savary}\ and\ \citenamefont
  {Balents}(2016)}]{sava:16:RPP}%
  \BibitemOpen
  \bibfield  {author} {\bibinfo {author} {\bibfnamefont {L.}~\bibnamefont
  {Savary}}\ and\ \bibinfo {author} {\bibfnamefont {L.}~\bibnamefont
  {Balents}},\ }\bibfield  {title} {\bibinfo {title} {Quantum spin liquids: a
  review},\ }\href {https://doi.org/10.1088/0034-4885/80/1/016502} {\bibfield
  {journal} {\bibinfo  {journal} {Reports on Progress in Physics}\ }\textbf
  {\bibinfo {volume} {80}},\ \bibinfo {pages} {016502} (\bibinfo {year}
  {2016})}\BibitemShut {NoStop}%
\bibitem [{\citenamefont {Balents}(2010)}]{bale:10:nature}%
  \BibitemOpen
  \bibfield  {author} {\bibinfo {author} {\bibfnamefont {L.}~\bibnamefont
  {Balents}},\ }\bibfield  {title} {\bibinfo {title} {Spin liquids in
  frustrated magnets},\ }\href {https://doi.org/10.1038/nature08917} {\bibfield
   {journal} {\bibinfo  {journal} {Nature}\ }\textbf {\bibinfo {volume}
  {464}},\ \bibinfo {pages} {199} (\bibinfo {year} {2010})}\BibitemShut
  {NoStop}%
\bibitem [{\citenamefont {Fishman}\ \emph {et~al.}(2018)\citenamefont
  {Fishman}, \citenamefont {Fernandez-Baca},\ and\ \citenamefont
  {Rõõm}}]{fish:18:book}%
  \BibitemOpen
  \bibfield  {author} {\bibinfo {author} {\bibfnamefont {R.~S.}\ \bibnamefont
  {Fishman}}, \bibinfo {author} {\bibfnamefont {J.~A.}\ \bibnamefont
  {Fernandez-Baca}},\ and\ \bibinfo {author} {\bibfnamefont {T.}~\bibnamefont
  {Rõõm}},\ }\href {https://doi.org/10.1088/978-1-64327-114-9} {\emph
  {\bibinfo {title} {Spin-Wave Theory and its Applications to Neutron
  Scattering and THz Spectroscopy}}},\ 2053-2571\ (\bibinfo  {publisher}
  {Morgan \& Claypool Publishers},\ \bibinfo {year} {2018})\BibitemShut
  {NoStop}%
\bibitem [{\citenamefont {Dieny}\ \emph {et~al.}(2011)\citenamefont {Dieny},
  \citenamefont {Sousa}, \citenamefont {Hérault}, \citenamefont {Papusoi},
  \citenamefont {Prenat}, \citenamefont {Ebels}, \citenamefont {Houssameddine},
  \citenamefont {Rodmacq}, \citenamefont {Auffret}, \citenamefont
  {Prejbeanu-Buda}, \citenamefont {Cyrille}, \citenamefont {Delaet},
  \citenamefont {Redon}, \citenamefont {Ducruet}, \citenamefont {Nozieres},\
  and\ \citenamefont {Prejbeanu}}]{dien:11:HMM}%
  \BibitemOpen
  \bibfield  {author} {\bibinfo {author} {\bibfnamefont {B.}~\bibnamefont
  {Dieny}}, \bibinfo {author} {\bibfnamefont {R.}~\bibnamefont {Sousa}},
  \bibinfo {author} {\bibfnamefont {J.}~\bibnamefont {Hérault}}, \bibinfo
  {author} {\bibfnamefont {C.}~\bibnamefont {Papusoi}}, \bibinfo {author}
  {\bibfnamefont {G.}~\bibnamefont {Prenat}}, \bibinfo {author} {\bibfnamefont
  {U.}~\bibnamefont {Ebels}}, \bibinfo {author} {\bibfnamefont
  {D.}~\bibnamefont {Houssameddine}}, \bibinfo {author} {\bibfnamefont
  {B.}~\bibnamefont {Rodmacq}}, \bibinfo {author} {\bibfnamefont
  {S.}~\bibnamefont {Auffret}}, \bibinfo {author} {\bibfnamefont
  {L.}~\bibnamefont {Prejbeanu-Buda}}, \bibinfo {author} {\bibfnamefont
  {M.}~\bibnamefont {Cyrille}}, \bibinfo {author} {\bibfnamefont
  {B.}~\bibnamefont {Delaet}}, \bibinfo {author} {\bibfnamefont
  {O.}~\bibnamefont {Redon}}, \bibinfo {author} {\bibfnamefont
  {C.}~\bibnamefont {Ducruet}}, \bibinfo {author} {\bibfnamefont
  {J.}~\bibnamefont {Nozieres}},\ and\ \bibinfo {author} {\bibfnamefont
  {L.}~\bibnamefont {Prejbeanu}},\ }\bibfield  {title} {\bibinfo {title}
  {Chapter two - spintronic devices for memory and logic applications},\ }\href
  {https://doi.org/https://doi.org/10.1016/B978-0-444-53780-5.00002-8}
  {\bibfield  {journal} {\bibinfo  {journal} {Handbook of Magnetic Materials}\
  }\textbf {\bibinfo {volume} {19}},\ \bibinfo {pages} {107 } (\bibinfo {year}
  {2011})}\BibitemShut {NoStop}%
\bibitem [{\citenamefont {Essafi}\ \emph {et~al.}(2017)\citenamefont {Essafi},
  \citenamefont {Benton},\ and\ \citenamefont {Jaubert}}]{essa:17:PRB}%
  \BibitemOpen
  \bibfield  {author} {\bibinfo {author} {\bibfnamefont {K.}~\bibnamefont
  {Essafi}}, \bibinfo {author} {\bibfnamefont {O.}~\bibnamefont {Benton}},\
  and\ \bibinfo {author} {\bibfnamefont {L.~D.~C.}\ \bibnamefont {Jaubert}},\
  }\bibfield  {title} {\bibinfo {title} {{Generic nearest-neighbor kagome
  model: XYZ and Dzyaloshinskii-Moriya couplings with comparison to the
  pyrochlore-lattice case}},\ }\href
  {https://doi.org/10.1103/PhysRevB.96.205126} {\bibfield  {journal} {\bibinfo
  {journal} {Phys. Rev. B}\ }\textbf {\bibinfo {volume} {96}},\ \bibinfo
  {pages} {205126} (\bibinfo {year} {2017})}\BibitemShut {NoStop}%
\bibitem [{\citenamefont {Ross}\ \emph {et~al.}(2009)\citenamefont {Ross},
  \citenamefont {Ruff}, \citenamefont {Adams}, \citenamefont {Gardner},
  \citenamefont {Dabkowska}, \citenamefont {Qiu}, \citenamefont {Copley},\ and\
  \citenamefont {Gaulin}}]{ross:09:PRL}%
  \BibitemOpen
  \bibfield  {author} {\bibinfo {author} {\bibfnamefont {K.~A.}\ \bibnamefont
  {Ross}}, \bibinfo {author} {\bibfnamefont {J.~P.~C.}\ \bibnamefont {Ruff}},
  \bibinfo {author} {\bibfnamefont {C.~P.}\ \bibnamefont {Adams}}, \bibinfo
  {author} {\bibfnamefont {J.~S.}\ \bibnamefont {Gardner}}, \bibinfo {author}
  {\bibfnamefont {H.~A.}\ \bibnamefont {Dabkowska}}, \bibinfo {author}
  {\bibfnamefont {Y.}~\bibnamefont {Qiu}}, \bibinfo {author} {\bibfnamefont
  {J.~R.~D.}\ \bibnamefont {Copley}},\ and\ \bibinfo {author} {\bibfnamefont
  {B.~D.}\ \bibnamefont {Gaulin}},\ }\bibfield  {title} {\bibinfo {title}
  {{Two-Dimensional Kagome Correlations and Field Induced Order in the
  Ferromagnetic $XY$ Pyrochlore
  ${\mathrm{Yb}}_{2}{\mathrm{Ti}}_{2}{\mathbf{O}}_{7}$}},\ }\href
  {https://doi.org/10.1103/PhysRevLett.103.227202} {\bibfield  {journal}
  {\bibinfo  {journal} {Phys. Rev. Lett.}\ }\textbf {\bibinfo {volume} {103}},\
  \bibinfo {pages} {227202} (\bibinfo {year} {2009})}\BibitemShut {NoStop}%
\bibitem [{\citenamefont {Gao}\ \emph {et~al.}(2018)\citenamefont {Gao},
  \citenamefont {Guratinder}, \citenamefont {Stuhr}, \citenamefont {White},
  \citenamefont {Mansson}, \citenamefont {Roessli}, \citenamefont {Fennell},
  \citenamefont {Tsurkan}, \citenamefont {Loidl}, \citenamefont
  {Ciomaga~Hatnean}, \citenamefont {Balakrishnan}, \citenamefont {Raymond},
  \citenamefont {Chapon}, \citenamefont {Garlea}, \citenamefont {Savici},
  \citenamefont {Cervellino}, \citenamefont {Bombardi}, \citenamefont
  {Chernyshov}, \citenamefont {R\"uegg}, \citenamefont {Haraldsen},\ and\
  \citenamefont {Zaharko}}]{gao:18:PRB}%
  \BibitemOpen
  \bibfield  {author} {\bibinfo {author} {\bibfnamefont {S.}~\bibnamefont
  {Gao}}, \bibinfo {author} {\bibfnamefont {K.}~\bibnamefont {Guratinder}},
  \bibinfo {author} {\bibfnamefont {U.}~\bibnamefont {Stuhr}}, \bibinfo
  {author} {\bibfnamefont {J.~S.}\ \bibnamefont {White}}, \bibinfo {author}
  {\bibfnamefont {M.}~\bibnamefont {Mansson}}, \bibinfo {author} {\bibfnamefont
  {B.}~\bibnamefont {Roessli}}, \bibinfo {author} {\bibfnamefont
  {T.}~\bibnamefont {Fennell}}, \bibinfo {author} {\bibfnamefont
  {V.}~\bibnamefont {Tsurkan}}, \bibinfo {author} {\bibfnamefont
  {A.}~\bibnamefont {Loidl}}, \bibinfo {author} {\bibfnamefont
  {M.}~\bibnamefont {Ciomaga~Hatnean}}, \bibinfo {author} {\bibfnamefont
  {G.}~\bibnamefont {Balakrishnan}}, \bibinfo {author} {\bibfnamefont
  {S.}~\bibnamefont {Raymond}}, \bibinfo {author} {\bibfnamefont
  {L.}~\bibnamefont {Chapon}}, \bibinfo {author} {\bibfnamefont {V.~O.}\
  \bibnamefont {Garlea}}, \bibinfo {author} {\bibfnamefont {A.~T.}\
  \bibnamefont {Savici}}, \bibinfo {author} {\bibfnamefont {A.}~\bibnamefont
  {Cervellino}}, \bibinfo {author} {\bibfnamefont {A.}~\bibnamefont
  {Bombardi}}, \bibinfo {author} {\bibfnamefont {D.}~\bibnamefont
  {Chernyshov}}, \bibinfo {author} {\bibfnamefont {C.}~\bibnamefont {R\"uegg}},
  \bibinfo {author} {\bibfnamefont {J.~T.}\ \bibnamefont {Haraldsen}},\ and\
  \bibinfo {author} {\bibfnamefont {O.}~\bibnamefont {Zaharko}},\ }\bibfield
  {title} {\bibinfo {title} {{Manifolds of magnetic ordered states and
  excitations in the almost Heisenberg pyrochlore antiferromagnet
  ${\mathrm{MgCr}}_{2}{\mathrm{O}}_{4}$}},\ }\href
  {https://doi.org/10.1103/PhysRevB.97.134430} {\bibfield  {journal} {\bibinfo
  {journal} {Phys. Rev. B}\ }\textbf {\bibinfo {volume} {97}},\ \bibinfo
  {pages} {134430} (\bibinfo {year} {2018})}\BibitemShut {NoStop}%
\bibitem [{\citenamefont {Benton}(2020)}]{bent:20:arx}%
  \BibitemOpen
  \bibfield  {author} {\bibinfo {author} {\bibfnamefont {O.}~\bibnamefont
  {Benton}},\ }\bibfield  {title} {\bibinfo {title} {Ordered ground states of
  kagome magnets with generic exchange anisotropy},\ }\href@noop {} {\bibfield
  {journal} {\bibinfo  {journal} {arXiv: Strongly Correlated Electrons}\ }
  (\bibinfo {year} {2020})},\ \Eprint {https://arxiv.org/abs/2008.04677}
  {arXiv:2008.04677 [cond-mat.str-el]} \BibitemShut {NoStop}%
\bibitem [{\citenamefont {Boyko}\ \emph {et~al.}(2018)\citenamefont {Boyko},
  \citenamefont {Balatsky},\ and\ \citenamefont {Haraldsen}}]{boyk:18:PRB}%
  \BibitemOpen
  \bibfield  {author} {\bibinfo {author} {\bibfnamefont {D.}~\bibnamefont
  {Boyko}}, \bibinfo {author} {\bibfnamefont {A.~V.}\ \bibnamefont
  {Balatsky}},\ and\ \bibinfo {author} {\bibfnamefont {J.~T.}\ \bibnamefont
  {Haraldsen}},\ }\bibfield  {title} {\bibinfo {title} {{Evolution of magnetic
  Dirac bosons in a honeycomb lattice}},\ }\href
  {https://doi.org/10.1103/PhysRevB.97.014433} {\bibfield  {journal} {\bibinfo
  {journal} {Phys. Rev. B}\ }\textbf {\bibinfo {volume} {97}},\ \bibinfo
  {pages} {014433} (\bibinfo {year} {2018})}\BibitemShut {NoStop}%
\bibitem [{\citenamefont {Wang}\ \emph {et~al.}(2020)\citenamefont {Wang},
  \citenamefont {Jiang}, \citenamefont {Hu}, \citenamefont {Wang},
  \citenamefont {Zhou}, \citenamefont {Yuan},\ and\ \citenamefont
  {Zhao}}]{wang:20:PCCP}%
  \BibitemOpen
  \bibfield  {author} {\bibinfo {author} {\bibfnamefont {P.}~\bibnamefont
  {Wang}}, \bibinfo {author} {\bibfnamefont {X.}~\bibnamefont {Jiang}},
  \bibinfo {author} {\bibfnamefont {J.}~\bibnamefont {Hu}}, \bibinfo {author}
  {\bibfnamefont {B.}~\bibnamefont {Wang}}, \bibinfo {author} {\bibfnamefont
  {T.}~\bibnamefont {Zhou}}, \bibinfo {author} {\bibfnamefont {H.}~\bibnamefont
  {Yuan}},\ and\ \bibinfo {author} {\bibfnamefont {J.}~\bibnamefont {Zhao}},\
  }\bibfield  {title} {\bibinfo {title} {Robust spin manipulation in 2d
  organometallic kagome lattices: a first-principles study},\ }\href
  {https://doi.org/10.1039/d0cp00742k} {\bibfield  {journal} {\bibinfo
  {journal} {Physical chemistry chemical physics : PCCP}\ }\textbf {\bibinfo
  {volume} {22}},\ \bibinfo {pages} {11045—11052} (\bibinfo {year}
  {2020})}\BibitemShut {NoStop}%
\bibitem [{\citenamefont {Harris}\ and\ \citenamefont
  {Yildirim}(2013)}]{harr:13:PRB}%
  \BibitemOpen
  \bibfield  {author} {\bibinfo {author} {\bibfnamefont {A.~B.}\ \bibnamefont
  {Harris}}\ and\ \bibinfo {author} {\bibfnamefont {T.}~\bibnamefont
  {Yildirim}},\ }\bibfield  {title} {\bibinfo {title} {{Spin dynamics of
  trimers on a distorted Kagome lattice}},\ }\href
  {https://doi.org/10.1103/PhysRevB.88.014411} {\bibfield  {journal} {\bibinfo
  {journal} {Phys. Rev. B}\ }\textbf {\bibinfo {volume} {88}},\ \bibinfo
  {pages} {014411} (\bibinfo {year} {2013})}\BibitemShut {NoStop}%
\bibitem [{\citenamefont {O'Brien}\ \emph {et~al.}(2010)\citenamefont
  {O'Brien}, \citenamefont {Pollmann},\ and\ \citenamefont
  {Fulde}}]{obri:10:PRB}%
  \BibitemOpen
  \bibfield  {author} {\bibinfo {author} {\bibfnamefont {A.}~\bibnamefont
  {O'Brien}}, \bibinfo {author} {\bibfnamefont {F.}~\bibnamefont {Pollmann}},\
  and\ \bibinfo {author} {\bibfnamefont {P.}~\bibnamefont {Fulde}},\ }\bibfield
   {title} {\bibinfo {title} {{Strongly correlated fermions on a Kagome
  lattice}},\ }\href {https://doi.org/10.1103/PhysRevB.81.235115} {\bibfield
  {journal} {\bibinfo  {journal} {Phys. Rev. B}\ }\textbf {\bibinfo {volume}
  {81}},\ \bibinfo {pages} {235115} (\bibinfo {year} {2010})}\BibitemShut
  {NoStop}%
\bibitem [{\citenamefont {Li}\ \emph {et~al.}(2018)\citenamefont {Li},
  \citenamefont {Zhuang}, \citenamefont {Wang}, \citenamefont {Feng},
  \citenamefont {Gao}, \citenamefont {Xu}, \citenamefont {Hao}, \citenamefont
  {Wang}, \citenamefont {Zhang}, \citenamefont {Wu}, \citenamefont {Dou},
  \citenamefont {Chen}, \citenamefont {Hu},\ and\ \citenamefont
  {Du}}]{li:18:SA}%
  \BibitemOpen
  \bibfield  {author} {\bibinfo {author} {\bibfnamefont {Z.}~\bibnamefont
  {Li}}, \bibinfo {author} {\bibfnamefont {J.}~\bibnamefont {Zhuang}}, \bibinfo
  {author} {\bibfnamefont {L.}~\bibnamefont {Wang}}, \bibinfo {author}
  {\bibfnamefont {H.}~\bibnamefont {Feng}}, \bibinfo {author} {\bibfnamefont
  {Q.}~\bibnamefont {Gao}}, \bibinfo {author} {\bibfnamefont {X.}~\bibnamefont
  {Xu}}, \bibinfo {author} {\bibfnamefont {W.}~\bibnamefont {Hao}}, \bibinfo
  {author} {\bibfnamefont {X.}~\bibnamefont {Wang}}, \bibinfo {author}
  {\bibfnamefont {C.}~\bibnamefont {Zhang}}, \bibinfo {author} {\bibfnamefont
  {K.}~\bibnamefont {Wu}}, \bibinfo {author} {\bibfnamefont {S.~X.}\
  \bibnamefont {Dou}}, \bibinfo {author} {\bibfnamefont {L.}~\bibnamefont
  {Chen}}, \bibinfo {author} {\bibfnamefont {Z.}~\bibnamefont {Hu}},\ and\
  \bibinfo {author} {\bibfnamefont {Y.}~\bibnamefont {Du}},\ }\bibfield
  {title} {\bibinfo {title} {{Realization of flat band with possible nontrivial
  topology in electronic Kagome lattice}},\ }\bibfield  {journal} {\bibinfo
  {journal} {Science Advances}\ }\textbf {\bibinfo {volume} {4}},\ \href
  {https://doi.org/10.1126/sciadv.aau4511} {10.1126/sciadv.aau4511} (\bibinfo
  {year} {2018})\BibitemShut {NoStop}%
\bibitem [{\citenamefont {Tran}\ \emph {et~al.}(2020)\citenamefont {Tran},
  \citenamefont {Nguyen}, \citenamefont {Nguyen},\ and\ \citenamefont
  {Tran}}]{tran:20:arx}%
  \BibitemOpen
  \bibfield  {author} {\bibinfo {author} {\bibfnamefont {M.}~\bibnamefont
  {Tran}}, \bibinfo {author} {\bibfnamefont {D.-B.}\ \bibnamefont {Nguyen}},
  \bibinfo {author} {\bibfnamefont {H.}~\bibnamefont {Nguyen}},\ and\ \bibinfo
  {author} {\bibfnamefont {T.~T.}\ \bibnamefont {Tran}},\ }\bibfield  {title}
  {\bibinfo {title} {Magnetic competition in topological kagome magnets},\
  }\href@noop {} {\bibfield  {journal} {\bibinfo  {journal} {arXiv: Strongly
  Correlated Electrons}\ } (\bibinfo {year} {2020})}\BibitemShut {NoStop}%
\bibitem [{\citenamefont {Maksymenko}\ \emph {et~al.}(2017)\citenamefont
  {Maksymenko}, \citenamefont {Moessner},\ and\ \citenamefont
  {Shtengel}}]{maks:17:PRB}%
  \BibitemOpen
  \bibfield  {author} {\bibinfo {author} {\bibfnamefont {M.}~\bibnamefont
  {Maksymenko}}, \bibinfo {author} {\bibfnamefont {R.}~\bibnamefont
  {Moessner}},\ and\ \bibinfo {author} {\bibfnamefont {K.}~\bibnamefont
  {Shtengel}},\ }\bibfield  {title} {\bibinfo {title} {{Persistence of the flat
  band in a Kagome magnet with dipolar interactions}},\ }\href
  {https://doi.org/10.1103/PhysRevB.96.134411} {\bibfield  {journal} {\bibinfo
  {journal} {Phys. Rev. B}\ }\textbf {\bibinfo {volume} {96}},\ \bibinfo
  {pages} {134411} (\bibinfo {year} {2017})}\BibitemShut {NoStop}%
\bibitem [{\citenamefont {Ochiai}\ \emph {et~al.}(2017)\citenamefont {Ochiai},
  \citenamefont {Seki},\ and\ \citenamefont {Okunishi}}]{ochi:17:JPSJ}%
  \BibitemOpen
  \bibfield  {author} {\bibinfo {author} {\bibfnamefont {M.}~\bibnamefont
  {Ochiai}}, \bibinfo {author} {\bibfnamefont {K.}~\bibnamefont {Seki}},\ and\
  \bibinfo {author} {\bibfnamefont {K.}~\bibnamefont {Okunishi}},\ }\bibfield
  {title} {\bibinfo {title} {{Spin-Wave Analysis for Kagome-Triangular Spin
  System and Coupled Spin Tubes: Low-Energy Excitation for the Cuboc Order}},\
  }\href {https://doi.org/10.7566/JPSJ.86.114701} {\bibfield  {journal}
  {\bibinfo  {journal} {Journal of the Physical Society of Japan}\ }\textbf
  {\bibinfo {volume} {86}},\ \bibinfo {pages} {114701} (\bibinfo {year}
  {2017})}\BibitemShut {NoStop}%
\bibitem [{\citenamefont {Schmalfuss}\ \emph {et~al.}(2004)\citenamefont
  {Schmalfuss}, \citenamefont {Richter},\ and\ \citenamefont
  {Ihle}}]{schm:04:PRB}%
  \BibitemOpen
  \bibfield  {author} {\bibinfo {author} {\bibfnamefont {D.}~\bibnamefont
  {Schmalfuss}}, \bibinfo {author} {\bibfnamefont {J.}~\bibnamefont
  {Richter}},\ and\ \bibinfo {author} {\bibfnamefont {D.}~\bibnamefont
  {Ihle}},\ }\bibfield  {title} {\bibinfo {title} {{Absence of long-range order
  in a spin-half Heisenberg antiferromagnet on the stacked Kagome lattice}},\
  }\bibfield  {journal} {\bibinfo  {journal} {Physical Review B}\ }\textbf
  {\bibinfo {volume} {70}},\ \href {https://doi.org/10.1103/PhysRevB.70.184412}
  {10.1103/PhysRevB.70.184412} (\bibinfo {year} {2004})\BibitemShut {NoStop}%
\bibitem [{\citenamefont {Maksymenko}\ \emph {et~al.}(2015)\citenamefont
  {Maksymenko}, \citenamefont {Chandra},\ and\ \citenamefont
  {Moessner}}]{maks:15:PRB}%
  \BibitemOpen
  \bibfield  {author} {\bibinfo {author} {\bibfnamefont {M.}~\bibnamefont
  {Maksymenko}}, \bibinfo {author} {\bibfnamefont {V.~R.}\ \bibnamefont
  {Chandra}},\ and\ \bibinfo {author} {\bibfnamefont {R.}~\bibnamefont
  {Moessner}},\ }\bibfield  {title} {\bibinfo {title} {{Classical dipoles on
  the Kagome lattice}},\ }\href {https://doi.org/10.1103/PhysRevB.91.184407}
  {\bibfield  {journal} {\bibinfo  {journal} {Phys. Rev. B}\ }\textbf {\bibinfo
  {volume} {91}},\ \bibinfo {pages} {184407} (\bibinfo {year}
  {2015})}\BibitemShut {NoStop}%
\bibitem [{\citenamefont {Yamanaka}\ \emph {et~al.}(2004)\citenamefont
  {Yamanaka}, \citenamefont {Yamaki}, \citenamefont {Nagao},\ and\
  \citenamefont {Yamaguchi}}]{yama:04:IJQC}%
  \BibitemOpen
  \bibfield  {author} {\bibinfo {author} {\bibfnamefont {S.}~\bibnamefont
  {Yamanaka}}, \bibinfo {author} {\bibfnamefont {D.}~\bibnamefont {Yamaki}},
  \bibinfo {author} {\bibfnamefont {H.}~\bibnamefont {Nagao}},\ and\ \bibinfo
  {author} {\bibfnamefont {K.}~\bibnamefont {Yamaguchi}},\ }\bibfield  {title}
  {\bibinfo {title} {{J‐model for magnetism and superconductivity of
  triangular, kagome, and related spin lattice systems}},\ }\href
  {https://doi.org/10.1002/qua.20299} {\bibfield  {journal} {\bibinfo
  {journal} {International Journal of Quantum Chemistry}\ }\textbf {\bibinfo
  {volume} {100}},\ \bibinfo {pages} {1179 } (\bibinfo {year}
  {2004})}\BibitemShut {NoStop}%
\bibitem [{\citenamefont {Moessner}(2000)}]{moes:00:CJP}%
  \BibitemOpen
  \bibfield  {author} {\bibinfo {author} {\bibfnamefont {R.}~\bibnamefont
  {Moessner}},\ }\bibfield  {title} {\bibinfo {title} {Magnets with strong
  geometric frustration},\ }\bibfield  {journal} {\bibinfo  {journal} {Canadian
  Journal of Physics}\ }\textbf {\bibinfo {volume} {79}},\ \href
  {https://doi.org/10.1139/cjp-79-11-12-1283} {10.1139/cjp-79-11-12-1283}
  (\bibinfo {year} {2000})\BibitemShut {NoStop}%
\bibitem [{\citenamefont {Gov}(2000)}]{gov:02:JPCM}%
  \BibitemOpen
  \bibfield  {author} {\bibinfo {author} {\bibfnamefont {N.}~\bibnamefont
  {Gov}},\ }\bibfield  {title} {\bibinfo {title} {{Coherent dipolar
  correlations in the ground-state of Kagome frustrated antiferromagnets}},\
  }\href@noop {} {\bibfield  {journal} {\bibinfo  {journal} {arXiv: Condensed
  Matter}\ } (\bibinfo {year} {2000})}\BibitemShut {NoStop}%
\bibitem [{\citenamefont {Owerre}\ \emph {et~al.}(2016)\citenamefont {Owerre},
  \citenamefont {Burkov},\ and\ \citenamefont {Melko}}]{ower:16:PRB}%
  \BibitemOpen
  \bibfield  {author} {\bibinfo {author} {\bibfnamefont {S.~A.}\ \bibnamefont
  {Owerre}}, \bibinfo {author} {\bibfnamefont {A.~A.}\ \bibnamefont {Burkov}},\
  and\ \bibinfo {author} {\bibfnamefont {R.~G.}\ \bibnamefont {Melko}},\
  }\bibfield  {title} {\bibinfo {title} {Linear spin-wave study of a quantum
  kagome ice},\ }\href {https://doi.org/10.1103/PhysRevB.93.144402} {\bibfield
  {journal} {\bibinfo  {journal} {Phys. Rev. B}\ }\textbf {\bibinfo {volume}
  {93}},\ \bibinfo {pages} {144402} (\bibinfo {year} {2016})}\BibitemShut
  {NoStop}%
\bibitem [{\citenamefont {Zhang}\ \emph {et~al.}(2020)\citenamefont {Zhang},
  \citenamefont {Feng}, \citenamefont {Heitmann}, \citenamefont {Kolesnikov},
  \citenamefont {Stone}, \citenamefont {Lu},\ and\ \citenamefont
  {Ke}}]{zhan:20:PRB}%
  \BibitemOpen
  \bibfield  {author} {\bibinfo {author} {\bibfnamefont {H.}~\bibnamefont
  {Zhang}}, \bibinfo {author} {\bibfnamefont {X.}~\bibnamefont {Feng}},
  \bibinfo {author} {\bibfnamefont {T.}~\bibnamefont {Heitmann}}, \bibinfo
  {author} {\bibfnamefont {A.~I.}\ \bibnamefont {Kolesnikov}}, \bibinfo
  {author} {\bibfnamefont {M.~B.}\ \bibnamefont {Stone}}, \bibinfo {author}
  {\bibfnamefont {Y.-M.}\ \bibnamefont {Lu}},\ and\ \bibinfo {author}
  {\bibfnamefont {X.}~\bibnamefont {Ke}},\ }\bibfield  {title} {\bibinfo
  {title} {{Topological magnon bands in a room-temperature Kagome magnet}},\
  }\href {https://doi.org/10.1103/PhysRevB.101.100405} {\bibfield  {journal}
  {\bibinfo  {journal} {Phys. Rev. B}\ }\textbf {\bibinfo {volume} {101}},\
  \bibinfo {pages} {100405} (\bibinfo {year} {2020})}\BibitemShut {NoStop}%
\bibitem [{\citenamefont {Sharma}\ and\ \citenamefont
  {Parimi}(2020)}]{shar:20:IEEE}%
  \BibitemOpen
  \bibfield  {author} {\bibinfo {author} {\bibfnamefont {D.}~\bibnamefont
  {Sharma}}\ and\ \bibinfo {author} {\bibfnamefont {P.}~\bibnamefont
  {Parimi}},\ }\bibfield  {title} {\bibinfo {title} {{Talbot Effect at the
  Dirac-Like Cone in Kagome Lattice Microwave Photonic Crystal}},\ }\href
  {https://doi.org/10.1109/LMWC.2020.2990060} {\bibfield  {journal} {\bibinfo
  {journal} {IEEE Microwave and Wireless Components Letters}\ }\textbf
  {\bibinfo {volume} {PP}},\ \bibinfo {pages} {1} (\bibinfo {year}
  {2020})}\BibitemShut {NoStop}%
\bibitem [{\citenamefont {Wu}\ and\ \citenamefont {Guo}(2019)}]{wu:19:IEEE}%
  \BibitemOpen
  \bibfield  {author} {\bibinfo {author} {\bibfnamefont {T.}~\bibnamefont
  {Wu}}\ and\ \bibinfo {author} {\bibfnamefont {J.}~\bibnamefont {Guo}},\
  }\bibfield  {title} {\bibinfo {title} {{Performance Potential of 2D Kagome
  Lattice Interconnects}},\ }\href {https://doi.org/10.1109/LED.2019.2947285}
  {\bibfield  {journal} {\bibinfo  {journal} {IEEE Electron Device Letters}\
  }\textbf {\bibinfo {volume} {PP}},\ \bibinfo {pages} {1} (\bibinfo {year}
  {2019})}\BibitemShut {NoStop}%
\bibitem [{\citenamefont {Yin}\ \emph {et~al.}(2018)\citenamefont {Yin},
  \citenamefont {Zhang}, \citenamefont {Li}, \citenamefont {Jiang},
  \citenamefont {Chang}, \citenamefont {Zhang}, \citenamefont {Lian},
  \citenamefont {Xiang}, \citenamefont {Belopolski}, \citenamefont {Zheng},
  \citenamefont {Cochran}, \citenamefont {Xu}, \citenamefont {Bian},
  \citenamefont {Liu}, \citenamefont {Chang}, \citenamefont {Lin},
  \citenamefont {Lu}, \citenamefont {Wang}, \citenamefont {Jia},\ and\
  \citenamefont {Hasan}}]{yin:18:N}%
  \BibitemOpen
  \bibfield  {author} {\bibinfo {author} {\bibfnamefont {J.}~\bibnamefont
  {Yin}}, \bibinfo {author} {\bibfnamefont {S.}~\bibnamefont {Zhang}}, \bibinfo
  {author} {\bibfnamefont {H.}~\bibnamefont {Li}}, \bibinfo {author}
  {\bibfnamefont {K.}~\bibnamefont {Jiang}}, \bibinfo {author} {\bibfnamefont
  {G.}~\bibnamefont {Chang}}, \bibinfo {author} {\bibfnamefont
  {B.}~\bibnamefont {Zhang}}, \bibinfo {author} {\bibfnamefont
  {B.}~\bibnamefont {Lian}}, \bibinfo {author} {\bibfnamefont {C.}~\bibnamefont
  {Xiang}}, \bibinfo {author} {\bibfnamefont {I.}~\bibnamefont {Belopolski}},
  \bibinfo {author} {\bibfnamefont {H.}~\bibnamefont {Zheng}}, \bibinfo
  {author} {\bibfnamefont {T.}~\bibnamefont {Cochran}}, \bibinfo {author}
  {\bibfnamefont {S.-Y.}\ \bibnamefont {Xu}}, \bibinfo {author} {\bibfnamefont
  {G.}~\bibnamefont {Bian}}, \bibinfo {author} {\bibfnamefont {K.}~\bibnamefont
  {Liu}}, \bibinfo {author} {\bibfnamefont {T.-R.}\ \bibnamefont {Chang}},
  \bibinfo {author} {\bibfnamefont {H.}~\bibnamefont {Lin}}, \bibinfo {author}
  {\bibfnamefont {Z.-Y.}\ \bibnamefont {Lu}}, \bibinfo {author} {\bibfnamefont
  {Z.}~\bibnamefont {Wang}}, \bibinfo {author} {\bibfnamefont {S.}~\bibnamefont
  {Jia}},\ and\ \bibinfo {author} {\bibfnamefont {M.~Z.}\ \bibnamefont
  {Hasan}},\ }\bibfield  {title} {\bibinfo {title} {{Giant and anisotropic
  many-body spin–orbit tunability in a strongly correlated Kagome magnet}},\
  }\href {https://doi.org/10.1038/s41586-018-0502-7} {\bibfield  {journal}
  {\bibinfo  {journal} {Nature}\ }\textbf {\bibinfo {volume} {562}} (\bibinfo
  {year} {2018})}\BibitemShut {NoStop}%
\bibitem [{\citenamefont {Yin}\ \emph {et~al.}(2019)\citenamefont {Yin},
  \citenamefont {Zhang}, \citenamefont {Chang}, \citenamefont {Wang},
  \citenamefont {Tsirkin}, \citenamefont {Guguchia}, \citenamefont {Lian},
  \citenamefont {Zhou}, \citenamefont {Jiang}, \citenamefont {Belopolski},
  \citenamefont {Shumiya}, \citenamefont {Multer}, \citenamefont {Litskevich},
  \citenamefont {Cochran}, \citenamefont {Lin}, \citenamefont {Wang},
  \citenamefont {Neupert}, \citenamefont {Jia}, \citenamefont {Lei},\ and\
  \citenamefont {Hasan}}]{yin:19:NP}%
  \BibitemOpen
  \bibfield  {author} {\bibinfo {author} {\bibfnamefont {J.}~\bibnamefont
  {Yin}}, \bibinfo {author} {\bibfnamefont {S.}~\bibnamefont {Zhang}}, \bibinfo
  {author} {\bibfnamefont {G.}~\bibnamefont {Chang}}, \bibinfo {author}
  {\bibfnamefont {Q.}~\bibnamefont {Wang}}, \bibinfo {author} {\bibfnamefont
  {S.}~\bibnamefont {Tsirkin}}, \bibinfo {author} {\bibfnamefont
  {Z.}~\bibnamefont {Guguchia}}, \bibinfo {author} {\bibfnamefont
  {B.}~\bibnamefont {Lian}}, \bibinfo {author} {\bibfnamefont {H.}~\bibnamefont
  {Zhou}}, \bibinfo {author} {\bibfnamefont {K.}~\bibnamefont {Jiang}},
  \bibinfo {author} {\bibfnamefont {I.}~\bibnamefont {Belopolski}}, \bibinfo
  {author} {\bibfnamefont {N.}~\bibnamefont {Shumiya}}, \bibinfo {author}
  {\bibfnamefont {D.}~\bibnamefont {Multer}}, \bibinfo {author} {\bibfnamefont
  {M.}~\bibnamefont {Litskevich}}, \bibinfo {author} {\bibfnamefont
  {T.}~\bibnamefont {Cochran}}, \bibinfo {author} {\bibfnamefont
  {H.}~\bibnamefont {Lin}}, \bibinfo {author} {\bibfnamefont {Z.}~\bibnamefont
  {Wang}}, \bibinfo {author} {\bibfnamefont {T.}~\bibnamefont {Neupert}},
  \bibinfo {author} {\bibfnamefont {S.}~\bibnamefont {Jia}}, \bibinfo {author}
  {\bibfnamefont {H.}~\bibnamefont {Lei}},\ and\ \bibinfo {author}
  {\bibfnamefont {M.~Z.}\ \bibnamefont {Hasan}},\ }\bibfield  {title} {\bibinfo
  {title} {Negative flat band magnetism in a spin–orbit-coupled correlated
  kagome magnet},\ }\href {https://doi.org/10.1038/s41567-019-0426-7}
  {\bibfield  {journal} {\bibinfo  {journal} {Nature Physics}\ }\textbf
  {\bibinfo {volume} {15}} (\bibinfo {year} {2019})}\BibitemShut {NoStop}%
\bibitem [{\citenamefont {Yazyev}(2019)}]{yazy:19:NP}%
  \BibitemOpen
  \bibfield  {author} {\bibinfo {author} {\bibfnamefont {O.}~\bibnamefont
  {Yazyev}},\ }\bibfield  {title} {\bibinfo {title} {An upside-down magnet},\
  }\href {https://doi.org/10.1038/s41567-019-0451-6} {\bibfield  {journal}
  {\bibinfo  {journal} {Nature Physics}\ }\textbf {\bibinfo {volume} {15}}
  (\bibinfo {year} {2019})}\BibitemShut {NoStop}%
\bibitem [{\citenamefont {Rigol}\ and\ \citenamefont
  {Singh}(2007)}]{rigo:07:PRL}%
  \BibitemOpen
  \bibfield  {author} {\bibinfo {author} {\bibfnamefont {M.}~\bibnamefont
  {Rigol}}\ and\ \bibinfo {author} {\bibfnamefont {R.}~\bibnamefont {Singh}},\
  }\bibfield  {title} {\bibinfo {title} {{Magnetic Susceptibility of the Kagome
  Antiferromagnet ZnCu$_3$(OH)$_6$Cl$_2$}},\ }\href
  {https://doi.org/10.1103/PhysRevLett.98.207204} {\bibfield  {journal}
  {\bibinfo  {journal} {Physical review letters}\ }\textbf {\bibinfo {volume}
  {98}},\ \bibinfo {pages} {207204} (\bibinfo {year} {2007})}\BibitemShut
  {NoStop}%
\bibitem [{\citenamefont {Yin}\ \emph {et~al.}(2020)\citenamefont {Yin},
  \citenamefont {Shumiya}, \citenamefont {Jiang}, \citenamefont {Zhou},
  \citenamefont {Macam}, \citenamefont {Zhang}, \citenamefont {Sura},
  \citenamefont {Cheng}, \citenamefont {Guguchia}, \citenamefont {Li},
  \citenamefont {Wang}, \citenamefont {Litskevich}, \citenamefont {Belopolski},
  \citenamefont {Yang}, \citenamefont {Cochran}, \citenamefont {Chang},
  \citenamefont {Zhang}, \citenamefont {Andersen}, \citenamefont {Huang},\ and\
  \citenamefont {Hasan}}]{yin:20:NC}%
  \BibitemOpen
  \bibfield  {author} {\bibinfo {author} {\bibfnamefont {J.}~\bibnamefont
  {Yin}}, \bibinfo {author} {\bibfnamefont {N.}~\bibnamefont {Shumiya}},
  \bibinfo {author} {\bibfnamefont {Y.}~\bibnamefont {Jiang}}, \bibinfo
  {author} {\bibfnamefont {H.}~\bibnamefont {Zhou}}, \bibinfo {author}
  {\bibfnamefont {G.}~\bibnamefont {Macam}}, \bibinfo {author} {\bibfnamefont
  {S.}~\bibnamefont {Zhang}}, \bibinfo {author} {\bibfnamefont
  {H.}~\bibnamefont {Sura}}, \bibinfo {author} {\bibfnamefont {Z.}~\bibnamefont
  {Cheng}}, \bibinfo {author} {\bibfnamefont {Z.}~\bibnamefont {Guguchia}},
  \bibinfo {author} {\bibfnamefont {Y.}~\bibnamefont {Li}}, \bibinfo {author}
  {\bibfnamefont {Q.}~\bibnamefont {Wang}}, \bibinfo {author} {\bibfnamefont
  {M.}~\bibnamefont {Litskevich}}, \bibinfo {author} {\bibfnamefont
  {I.}~\bibnamefont {Belopolski}}, \bibinfo {author} {\bibfnamefont
  {X.}~\bibnamefont {Yang}}, \bibinfo {author} {\bibfnamefont {T.}~\bibnamefont
  {Cochran}}, \bibinfo {author} {\bibfnamefont {G.}~\bibnamefont {Chang}},
  \bibinfo {author} {\bibfnamefont {Q.}~\bibnamefont {Zhang}}, \bibinfo
  {author} {\bibfnamefont {B.}~\bibnamefont {Andersen}}, \bibinfo {author}
  {\bibfnamefont {Z.-Q.}\ \bibnamefont {Huang}},\ and\ \bibinfo {author}
  {\bibfnamefont {M.~Z.}\ \bibnamefont {Hasan}},\ }\bibfield  {title} {\bibinfo
  {title} {Spin-orbit quantum impurity in a topological kagome magnet},\
  }\href@noop {} {\bibfield  {journal} {\bibinfo  {journal} {Nature
  Communications}\ } (\bibinfo {year} {2020})}\BibitemShut {NoStop}%
\bibitem [{\citenamefont {Boyko}\ \emph {et~al.}(2020)\citenamefont {Boyko},
  \citenamefont {Saxena},\ and\ \citenamefont {Haraldsen}}]{boyk:20:AdP}%
  \BibitemOpen
  \bibfield  {author} {\bibinfo {author} {\bibfnamefont {D.}~\bibnamefont
  {Boyko}}, \bibinfo {author} {\bibfnamefont {A.}~\bibnamefont {Saxena}},\ and\
  \bibinfo {author} {\bibfnamefont {J.~T.}\ \bibnamefont {Haraldsen}},\
  }\bibfield  {title} {\bibinfo {title} {{Spin Dynamics and Dirac Nodes in a
  Kagome Lattice}},\ }\href {https://doi.org/10.1002/andp.201900350} {\bibfield
   {journal} {\bibinfo  {journal} {Annalen der Physik}\ }\textbf {\bibinfo
  {volume} {532}},\ \bibinfo {pages} {1900350} (\bibinfo {year}
  {2020})}\BibitemShut {NoStop}%
\bibitem [{\citenamefont {Han}\ \emph {et~al.}(2012)\citenamefont {Han},
  \citenamefont {Helton}, \citenamefont {Chu}, \citenamefont {Nocera},
  \citenamefont {Rodriguez-Rivera}, \citenamefont {Broholm},\ and\
  \citenamefont {Lee}}]{han:12:nature}%
  \BibitemOpen
  \bibfield  {author} {\bibinfo {author} {\bibfnamefont {T.-H.}\ \bibnamefont
  {Han}}, \bibinfo {author} {\bibfnamefont {J.~S.}\ \bibnamefont {Helton}},
  \bibinfo {author} {\bibfnamefont {S.}~\bibnamefont {Chu}}, \bibinfo {author}
  {\bibfnamefont {D.~G.}\ \bibnamefont {Nocera}}, \bibinfo {author}
  {\bibfnamefont {J.~A.}\ \bibnamefont {Rodriguez-Rivera}}, \bibinfo {author}
  {\bibfnamefont {C.}~\bibnamefont {Broholm}},\ and\ \bibinfo {author}
  {\bibfnamefont {Y.~S.}\ \bibnamefont {Lee}},\ }\bibfield  {title} {\bibinfo
  {title} {{Fractionalized excitations in the spin-liquid state of a
  Kagome-lattice antiferromagnet}},\ }\href
  {https://doi.org/10.1038/nature11659} {\bibfield  {journal} {\bibinfo
  {journal} {Nature}\ }\textbf {\bibinfo {volume} {492}},\ \bibinfo {pages}
  {406} (\bibinfo {year} {2012})}\BibitemShut {NoStop}%
\bibitem [{\citenamefont {Helton}\ \emph {et~al.}(2007)\citenamefont {Helton},
  \citenamefont {Matan}, \citenamefont {Shores}, \citenamefont {Nytko},
  \citenamefont {Bartlett}, \citenamefont {Yoshida}, \citenamefont {Takano},
  \citenamefont {Suslov}, \citenamefont {Qiu}, \citenamefont {Chung},
  \citenamefont {Nocera},\ and\ \citenamefont {Lee}}]{helt:07:PRL}%
  \BibitemOpen
  \bibfield  {author} {\bibinfo {author} {\bibfnamefont {J.~S.}\ \bibnamefont
  {Helton}}, \bibinfo {author} {\bibfnamefont {K.}~\bibnamefont {Matan}},
  \bibinfo {author} {\bibfnamefont {M.~P.}\ \bibnamefont {Shores}}, \bibinfo
  {author} {\bibfnamefont {E.~A.}\ \bibnamefont {Nytko}}, \bibinfo {author}
  {\bibfnamefont {B.~M.}\ \bibnamefont {Bartlett}}, \bibinfo {author}
  {\bibfnamefont {Y.}~\bibnamefont {Yoshida}}, \bibinfo {author} {\bibfnamefont
  {Y.}~\bibnamefont {Takano}}, \bibinfo {author} {\bibfnamefont
  {A.}~\bibnamefont {Suslov}}, \bibinfo {author} {\bibfnamefont
  {Y.}~\bibnamefont {Qiu}}, \bibinfo {author} {\bibfnamefont {J.-H.}\
  \bibnamefont {Chung}}, \bibinfo {author} {\bibfnamefont {D.~G.}\ \bibnamefont
  {Nocera}},\ and\ \bibinfo {author} {\bibfnamefont {Y.~S.}\ \bibnamefont
  {Lee}},\ }\bibfield  {title} {\bibinfo {title} {{Spin Dynamics of the
  Spin-$1/2$ Kagome Lattice Antiferromagnet
  ${\mathrm{ZnCu}}_{3}(\mathrm{OH}{)}_{6}{\mathrm{Cl}}_{2}$}},\ }\href
  {https://doi.org/10.1103/PhysRevLett.98.107204} {\bibfield  {journal}
  {\bibinfo  {journal} {Phys. Rev. Lett.}\ }\textbf {\bibinfo {volume} {98}},\
  \bibinfo {pages} {107204} (\bibinfo {year} {2007})}\BibitemShut {NoStop}%
\bibitem [{\citenamefont {Hirschberger}\ \emph {et~al.}(2015)\citenamefont
  {Hirschberger}, \citenamefont {Chisnell}, \citenamefont {Lee},\ and\
  \citenamefont {Ong}}]{hirs:15:PRL}%
  \BibitemOpen
  \bibfield  {author} {\bibinfo {author} {\bibfnamefont {M.}~\bibnamefont
  {Hirschberger}}, \bibinfo {author} {\bibfnamefont {R.}~\bibnamefont
  {Chisnell}}, \bibinfo {author} {\bibfnamefont {Y.~S.}\ \bibnamefont {Lee}},\
  and\ \bibinfo {author} {\bibfnamefont {N.~P.}\ \bibnamefont {Ong}},\
  }\bibfield  {title} {\bibinfo {title} {{Thermal Hall Effect of Spin
  Excitations in a Kagome Magnet}},\ }\href
  {https://doi.org/10.1103/PhysRevLett.115.106603} {\bibfield  {journal}
  {\bibinfo  {journal} {Phys. Rev. Lett.}\ }\textbf {\bibinfo {volume} {115}},\
  \bibinfo {pages} {106603} (\bibinfo {year} {2015})}\BibitemShut {NoStop}%
\bibitem [{\citenamefont {Dun}\ \emph {et~al.}(2016)\citenamefont {Dun},
  \citenamefont {Trinh}, \citenamefont {Li}, \citenamefont {Lee}, \citenamefont
  {Chen}, \citenamefont {Baumbach}, \citenamefont {Hu}, \citenamefont {Wang},
  \citenamefont {Choi}, \citenamefont {Shastry}, \citenamefont {Ramirez},\ and\
  \citenamefont {Zhou}}]{dun:16:PRL}%
  \BibitemOpen
  \bibfield  {author} {\bibinfo {author} {\bibfnamefont {Z.~L.}\ \bibnamefont
  {Dun}}, \bibinfo {author} {\bibfnamefont {J.}~\bibnamefont {Trinh}}, \bibinfo
  {author} {\bibfnamefont {K.}~\bibnamefont {Li}}, \bibinfo {author}
  {\bibfnamefont {M.}~\bibnamefont {Lee}}, \bibinfo {author} {\bibfnamefont
  {K.~W.}\ \bibnamefont {Chen}}, \bibinfo {author} {\bibfnamefont
  {R.}~\bibnamefont {Baumbach}}, \bibinfo {author} {\bibfnamefont {Y.~F.}\
  \bibnamefont {Hu}}, \bibinfo {author} {\bibfnamefont {Y.~X.}\ \bibnamefont
  {Wang}}, \bibinfo {author} {\bibfnamefont {E.~S.}\ \bibnamefont {Choi}},
  \bibinfo {author} {\bibfnamefont {B.~S.}\ \bibnamefont {Shastry}}, \bibinfo
  {author} {\bibfnamefont {A.~P.}\ \bibnamefont {Ramirez}},\ and\ \bibinfo
  {author} {\bibfnamefont {H.~D.}\ \bibnamefont {Zhou}},\ }\bibfield  {title}
  {\bibinfo {title} {{Magnetic Ground States of the Rare-Earth Tripod Kagome
  Lattice
  ${\mathrm{Mg}}_{2}{\mathrm{RE}}_{3}{\mathrm{Sb}}_{3}{\mathrm{O}}_{14}$
  ($\mathrm{RE}=\mathrm{Gd},\mathrm{Dy},\mathrm{Er}$)}},\ }\href
  {https://doi.org/10.1103/PhysRevLett.116.157201} {\bibfield  {journal}
  {\bibinfo  {journal} {Phys. Rev. Lett.}\ }\textbf {\bibinfo {volume} {116}},\
  \bibinfo {pages} {157201} (\bibinfo {year} {2016})}\BibitemShut {NoStop}%
\bibitem [{\citenamefont {Kassem}\ \emph {et~al.}(2017)\citenamefont {Kassem},
  \citenamefont {Tabata}, \citenamefont {Waki},\ and\ \citenamefont
  {Nakamura}}]{kass:17:PRB}%
  \BibitemOpen
  \bibfield  {author} {\bibinfo {author} {\bibfnamefont {M.~A.}\ \bibnamefont
  {Kassem}}, \bibinfo {author} {\bibfnamefont {Y.}~\bibnamefont {Tabata}},
  \bibinfo {author} {\bibfnamefont {T.}~\bibnamefont {Waki}},\ and\ \bibinfo
  {author} {\bibfnamefont {H.}~\bibnamefont {Nakamura}},\ }\bibfield  {title}
  {\bibinfo {title} {{Low-field anomalous magnetic phase in the Kagome-lattice
  shandite
  $\mathrm{C}{\mathrm{o}}_{3}\mathrm{S}{\mathrm{n}}_{2}{\mathrm{S}}_{2}$}},\
  }\href {https://doi.org/10.1103/PhysRevB.96.014429} {\bibfield  {journal}
  {\bibinfo  {journal} {Phys. Rev. B}\ }\textbf {\bibinfo {volume} {96}},\
  \bibinfo {pages} {014429} (\bibinfo {year} {2017})}\BibitemShut {NoStop}%
\bibitem [{\citenamefont {Zorko}\ \emph {et~al.}(2019)\citenamefont {Zorko},
  \citenamefont {Pregelj}, \citenamefont {Gomil\ifmmode~\check{s}\else
  \v{s}\fi{}ek}, \citenamefont {Klanj\ifmmode~\check{s}\else \v{s}\fi{}ek},
  \citenamefont {Zaharko}, \citenamefont {Sun},\ and\ \citenamefont
  {Mi}}]{zork:19:PRB}%
  \BibitemOpen
  \bibfield  {author} {\bibinfo {author} {\bibfnamefont {A.}~\bibnamefont
  {Zorko}}, \bibinfo {author} {\bibfnamefont {M.}~\bibnamefont {Pregelj}},
  \bibinfo {author} {\bibfnamefont {M.}~\bibnamefont
  {Gomil\ifmmode~\check{s}\else \v{s}\fi{}ek}}, \bibinfo {author}
  {\bibfnamefont {M.}~\bibnamefont {Klanj\ifmmode~\check{s}\else
  \v{s}\fi{}ek}}, \bibinfo {author} {\bibfnamefont {O.}~\bibnamefont
  {Zaharko}}, \bibinfo {author} {\bibfnamefont {W.}~\bibnamefont {Sun}},\ and\
  \bibinfo {author} {\bibfnamefont {J.-X.}\ \bibnamefont {Mi}},\ }\bibfield
  {title} {\bibinfo {title} {{Negative-vector-chirality
  ${120}^{\ensuremath{\circ}}$ spin structure in the defect- and
  distortion-free quantum Kagome antiferromagnet
  ${\mathrm{YCu}}_{3}{(\mathrm{OH})}_{6}{\mathrm{Cl}}_{3}$}},\ }\href
  {https://doi.org/10.1103/PhysRevB.100.144420} {\bibfield  {journal} {\bibinfo
   {journal} {Phys. Rev. B}\ }\textbf {\bibinfo {volume} {100}},\ \bibinfo
  {pages} {144420} (\bibinfo {year} {2019})}\BibitemShut {NoStop}%
\bibitem [{\citenamefont {Li}\ \emph {et~al.}(2010)\citenamefont {Li},
  \citenamefont {Gong}, \citenamefont {Zhao}, \citenamefont {Ran},
  \citenamefont {Gao},\ and\ \citenamefont {Su}}]{wei:10:PRB}%
  \BibitemOpen
  \bibfield  {author} {\bibinfo {author} {\bibfnamefont {W.}~\bibnamefont
  {Li}}, \bibinfo {author} {\bibfnamefont {S.-S.}\ \bibnamefont {Gong}},
  \bibinfo {author} {\bibfnamefont {Y.}~\bibnamefont {Zhao}}, \bibinfo {author}
  {\bibfnamefont {S.-J.}\ \bibnamefont {Ran}}, \bibinfo {author} {\bibfnamefont
  {S.}~\bibnamefont {Gao}},\ and\ \bibinfo {author} {\bibfnamefont
  {G.}~\bibnamefont {Su}},\ }\bibfield  {title} {\bibinfo {title} {{Phase
  transitions and thermodynamics of the two-dimensional Ising model on a
  distorted kagome lattice}},\ }\href
  {https://doi.org/10.1103/PhysRevB.82.134434} {\bibfield  {journal} {\bibinfo
  {journal} {Phys. Rev. B}\ }\textbf {\bibinfo {volume} {82}},\ \bibinfo
  {pages} {134434} (\bibinfo {year} {2010})}\BibitemShut {NoStop}%
\bibitem [{\citenamefont {Wulferding}\ \emph {et~al.}(2012)\citenamefont
  {Wulferding}, \citenamefont {Lemmens}, \citenamefont {Yoshida}, \citenamefont
  {Okamoto},\ and\ \citenamefont {Hiroi}}]{wulf:12:JPCM}%
  \BibitemOpen
  \bibfield  {author} {\bibinfo {author} {\bibfnamefont {D.}~\bibnamefont
  {Wulferding}}, \bibinfo {author} {\bibfnamefont {P.}~\bibnamefont {Lemmens}},
  \bibinfo {author} {\bibfnamefont {H.}~\bibnamefont {Yoshida}}, \bibinfo
  {author} {\bibfnamefont {Y.}~\bibnamefont {Okamoto}},\ and\ \bibinfo {author}
  {\bibfnamefont {Z.}~\bibnamefont {Hiroi}},\ }\bibfield  {title} {\bibinfo
  {title} {{The spin dynamics in distorted Kagome lattices: a comparative Raman
  study}},\ }\href {https://doi.org/10.1088/0953-8984/24/18/185602} {\bibfield
  {journal} {\bibinfo  {journal} {Journal of Physics: Condensed Matter}\
  }\textbf {\bibinfo {volume} {24}},\ \bibinfo {pages} {185602} (\bibinfo
  {year} {2012})}\BibitemShut {NoStop}%
\bibitem [{\citenamefont {Matan}\ \emph {et~al.}(2019)\citenamefont {Matan},
  \citenamefont {Ono}, \citenamefont {Gitgeatpong}, \citenamefont {de~Roos},
  \citenamefont {Miao}, \citenamefont {Torii}, \citenamefont {Kamiyama},
  \citenamefont {Miyata}, \citenamefont {Matsuo}, \citenamefont {Kindo},
  \citenamefont {Takeyama}, \citenamefont {Nambu}, \citenamefont
  {Piyawongwatthana}, \citenamefont {Sato},\ and\ \citenamefont
  {Tanaka}}]{mata:19:PRB}%
  \BibitemOpen
  \bibfield  {author} {\bibinfo {author} {\bibfnamefont {K.}~\bibnamefont
  {Matan}}, \bibinfo {author} {\bibfnamefont {T.}~\bibnamefont {Ono}}, \bibinfo
  {author} {\bibfnamefont {G.}~\bibnamefont {Gitgeatpong}}, \bibinfo {author}
  {\bibfnamefont {K.}~\bibnamefont {de~Roos}}, \bibinfo {author} {\bibfnamefont
  {P.}~\bibnamefont {Miao}}, \bibinfo {author} {\bibfnamefont {S.}~\bibnamefont
  {Torii}}, \bibinfo {author} {\bibfnamefont {T.}~\bibnamefont {Kamiyama}},
  \bibinfo {author} {\bibfnamefont {A.}~\bibnamefont {Miyata}}, \bibinfo
  {author} {\bibfnamefont {A.}~\bibnamefont {Matsuo}}, \bibinfo {author}
  {\bibfnamefont {K.}~\bibnamefont {Kindo}}, \bibinfo {author} {\bibfnamefont
  {S.}~\bibnamefont {Takeyama}}, \bibinfo {author} {\bibfnamefont
  {Y.}~\bibnamefont {Nambu}}, \bibinfo {author} {\bibfnamefont
  {P.}~\bibnamefont {Piyawongwatthana}}, \bibinfo {author} {\bibfnamefont
  {T.~J.}\ \bibnamefont {Sato}},\ and\ \bibinfo {author} {\bibfnamefont
  {H.}~\bibnamefont {Tanaka}},\ }\bibfield  {title} {\bibinfo {title}
  {{Magnetic structure and high-field magnetization of the distorted Kagome
  lattice antiferromagnet
  ${\mathrm{Cs}}_{2}{\mathrm{Cu}}_{3}{\mathrm{SnF}}_{12}$}},\ }\href
  {https://doi.org/10.1103/PhysRevB.99.224404} {\bibfield  {journal} {\bibinfo
  {journal} {Phys. Rev. B}\ }\textbf {\bibinfo {volume} {99}},\ \bibinfo
  {pages} {224404} (\bibinfo {year} {2019})}\BibitemShut {NoStop}%
\bibitem [{\citenamefont {Matan}\ \emph {et~al.}(2010)\citenamefont {Matan},
  \citenamefont {Ono}, \citenamefont {Fukumoto}, \citenamefont {Sato},
  \citenamefont {Yamaura}, \citenamefont {Yano}, \citenamefont {Morita},\ and\
  \citenamefont {Tanaka}}]{mata:10:NatPhys}%
  \BibitemOpen
  \bibfield  {author} {\bibinfo {author} {\bibfnamefont {K.}~\bibnamefont
  {Matan}}, \bibinfo {author} {\bibfnamefont {T.}~\bibnamefont {Ono}}, \bibinfo
  {author} {\bibfnamefont {Y.}~\bibnamefont {Fukumoto}}, \bibinfo {author}
  {\bibfnamefont {T.~J.}\ \bibnamefont {Sato}}, \bibinfo {author}
  {\bibfnamefont {J.}~\bibnamefont {Yamaura}}, \bibinfo {author} {\bibfnamefont
  {M.}~\bibnamefont {Yano}}, \bibinfo {author} {\bibfnamefont {K.}~\bibnamefont
  {Morita}},\ and\ \bibinfo {author} {\bibfnamefont {H.}~\bibnamefont
  {Tanaka}},\ }\bibfield  {title} {\bibinfo {title} {{Pinwheel valence-bond
  solid and triplet excitations in the two-dimensional deformed Kagome
  lattice}},\ }\href {https://doi.org/10.1038/nphys1761} {\bibfield  {journal}
  {\bibinfo  {journal} {Nature Physics}\ }\textbf {\bibinfo {volume} {6}},\
  \bibinfo {pages} {865} (\bibinfo {year} {2010})}\BibitemShut {NoStop}%
\end{thebibliography}%

\end{document}